# Structural Phase Transition and Interlayer Coupling in Few-Layer 1T′ and $T_d$ MoTe$_2$

*Yeryun Cheon, Soo Yeon Lim, Kangwon Kim and Hyeonsik Cheong\**

Department of Physics, Sogang University, Seoul 04107, Korea

\*E-mail: hcheong@sogang.ac.kr

ABSTRACT

We performed polarized Raman spectroscopy on mechanically exfoliated few-layer MoTe$_2$ samples and observed both 1T′ and $T_d$ phases at room temperature. Few-layer 1T′ and $T_d$ MoTe$_2$ exhibited a significant difference especially in interlayer vibration modes, from which the interlayer coupling strengths were extracted using the linear chain model: strong in-plane anisotropy was observed in both phases. Furthermore, temperature-dependent Raman measurements revealed a peculiar phase transition behavior in few-layer 1T′ MoTe$_2$. In contrast to bulk 1T′ MoTe$_2$ crystals where the phase transition to the $T_d$ phase occurs at ~250 K, the temperature-driven phase transition to the $T_d$ phase is increasingly suppressed as the thickness is reduced, and the transition and the critical temperature varied dramatically from sample to sample even for the same thickness. Raman spectra of intermediate phases that correspond to neither 1T′ nor $T_d$ phase with different interlayer vibration modes were observed, which suggests that several metastable phases exist with similar total energies.

TEXT

Molybdenum ditelluride (MoTe$_2$) exists in several phases even at room temperature with distinct physical properties.[1–4] At room temperature, both semiconducting hexagonal 2H phase and semimetallic monoclinic 1T′ phase are stable, and the phase transition between the two phases by electrostatic doping,[5,6] laser irradiation,[7,8] or strain[9,10] has been reported. By utilizing the phase transition, one can control the local electrical and topological properties and realize device structures that are not possible with a single-phase material.[11–13] Therefore, understanding the signatures of different phases and the transition between them is important. In the bulk case, the monoclinic 1T′ phase undergoes a temperature-driven structural phase transition to the orthorhombic $T_d$ phase at ~250 K,[14–17] whereas the 2H phase does not show any structural phase transition at low temperatures.[18] Figure 1a shows the crystal structure of individual monolayer MoTe$_2$ in both 1T′ and $T_d$ phases; the structure of the individual monolayer is identical in the two phases. Figures 1b and c compare the crystal structures of the 1T′ and $T_d$ phases: they differ only in the layer-to-layer stacking. The 1T′ phase is monoclinic with the stacking angle of ~93.9°, whereas the $T_d$ phase is orthorhombic. In the monoclinic stacking of the 1T′ phase, the symmetry elements are a two-fold screw axis along the $x$ direction ($C_{2x}$), a horizontal mirror plane ($M_x$), and a resulting inversion center, and thereby it belongs to the point group $C_{2h}$ ($2/m$), just like the individual monolayer. On the other hand, in the orthorhombic stacking of the $T_d$ phase, there are a two-fold screw axis along the $z$ direction ($C_{2z}$), a diagonal glide plane ($M_y$), and a vertical mirror plane ($M_x$), and so it belongs to the point group $C_{2v}$ ($mm2$). The $T_d$ phase lacks inversion symmetry which results in interesting quantum phenomena, including type-II Weyl semimetallic states,[19–24] the circular photogalvanic effect,[25,26] and the nonlinear Hall effect.[27–29]

For few-layer MoTe$_2$, on the other hand, controversy remains in the literature on the stable phase even at room temperature. Whereas there have been a few reports on few-layer 1T′ MoTe$_2$,[30–32] other reports claimed that MoTe$_2$ thin films below ~12 nm exist only in the $T_d$ phase at room temperature even though the samples were exfoliated from bulk 1T′ MoTe$_2$ crystals.[33] Few-layer 1T′ and $T_d$ MoTe$_2$ have different crystal symmetry from their bulk counterparts. The crystal symmetry of few-layer 1T′ MoTe$_2$ depends on the layer parity. Odd-layer 1T′ MoTe$_2$ shares the same symmetry elements and the point group $C_{2h}$ ($2/m$) with its bulk and monolayer forms, whereas even-layer 1T′ MoTe$_2$ does not hold a screw axis and thereby loses an inversion center, which leads to the point group $C_s$ ($m$) with only a mirror plane.[30] On the other hand, the crystal symmetry of few-layer $T_d$ MoTe$_2$ is independent of the number of layers, and only the mirror symmetry exists with the point group $C_s$ ($m$). The screw axis in the $z$ direction and the glide plane of its bulk counterpart no longer exist in few-layer $T_d$ MoTe$_2$, and so the symmetry analysis of few-layer 1T′ and $T_d$ MoTe$_2$ should in principle be different from that of bulk crystals. Furthermore, investigations on the temperature-dependence of few-layer MoTe$_2$ have not yet revealed clear indications of the 1T′ to $T_d$ phase transition.[32,34,35]

Raman spectroscopy is a powerful tool to investigate the crystal symmetry, interlayer coupling and layer stacking in two-dimensional materials.[36,37] In bulk crystals, the structural phase transition from the 1T′ to the $T_d$ phase leads to clear Raman signatures including the emergence of an interlayer shear mode,[16,17] but the interlayer vibration modes of few-layer 1T′ and $T_d$ MoTe$_2$ have not yet been studied in detail. In this work, we carried out polarized Raman measurements of exfoliated few-layer MoTe$_2$ samples and observed both 1T′ and $T_d$ phases at room temperature. Few-layer 1T′ and $T_d$ MoTe$_2$ showed unambiguous differences in interlayer vibration modes which were clarified based on symmetry analysis. We also extracted the interlayer coupling strength using the linear chain model and noticed strong in-plane anisotropy in both the phases. Furthermore, we performed temperature-dependent Raman spectroscopy on few-layer 1T′ MoTe$_2$ samples down to 10 K. The temperature-driven phase transition to the $T_d$ phase is increasingly suppressed as the thickness is reduced, and the transition and the critical temperature varied dramatically from sample to sample for the same thickness unlike the bulk crystals.

RESULTS AND DISCUSSION

Figure 1d compares the polarized Raman spectra of bulk MoTe$_2$ in two different phases: 1T′ phase at 295 K and $T_d$ phase at 78 K. The Raman spectrum of the low-temperature $T_d$ phase exhibits several differences from that of the room-temperature 1T′ phase: a low-frequency mode at 13 cm$^{-1}$ (S1) emerges, the peak at ~130 cm$^{-1}$ (P5) and ~190 cm$^{-1}$ (P7) are split into two peaks each. Because the two phases differ only in the way the layers are stacked, the frequencies of corresponding modes should be similar. On the other hand, the difference in the crystal symmetry should result in different selection rules in the two phases. Both 1T′ and $T_d$ phase contain the two layers with 12 atoms in the unit cell, and consequently each has 36 vibration modes in total. The vibration modes of the 1T′ phase reduce into $\Gamma = 12A_g + 6B_g + 6A_u + 12B_u$, where $A_g$ and $B_g$ modes are Raman-active, and $A_u$ and $B_u$ modes are Raman-inactive. On the other hand, the vibration modes of the $T_d$ phase decompose into $\Gamma = 12A_1 + 6A_2 + 6B_1 + 12B_2$, where all the modes are Raman-active, but only $A_1$ and $A_2$ modes can be observed in our backscattering geometry. Based on the Raman selection rules, the $A_g$ modes of the 1T′ phase and the $A_1$ modes of the $T_d$ phase are observed in the parallel polarization configuration along the crystalline axes [($xx$) or ($yy$)], whereas the $B_g$ modes of the 1T′ phase and the $A_2$ modes of the $T_d$ phase are observed in the cross polarization configuration along the crystalline axes [($xy$) or ($yx$)] (for more details, see Supporting Information Note S1). By comparing the polarization behaviors of the Raman peaks, one can assign P1, P4, P5, P6, P8 and P9 which are observed in the parallel polarization as the $A_g$ modes in the 1T′ phase and the $A_1$ modes in the $T_d$ phase and determine the crystalline axes (see Supporting Information Figure S1).[30,31,38] Also, P2, P3 and P7 observed in the cross polarization are assigned as the $B_g$ modes in the 1T′ phase and the $A_2$ modes in the $T_d$ phase. When the phase transition occurs, the $A_g$ and $B_u$ modes of the 1T′ phase become $A_1$ or $B_2$ modes in the $T_d$ phase, and the Raman peaks that appear in the $T_d$ phase at 13 cm$^{-1}$ (S1) and 128 cm$^{-1}$ (P5a) are the *Raman-active* $A_1$ modes corresponding to the *Raman-inactive* $B_u$ modes of the 1T′ phase. Similarly, the $B_g$ and $A_u$ modes of the 1T′ phase become $A_2$ or $B_1$ modes in the $T_d$ phase, and the Raman peak at 187 cm$^{-1}$ (P7a) in the $T_d$ phase is a *Raman-active* $A_2$ mode corresponding to the *Raman-inactive* $A_u$ mode of the 1T′ phase.[16,17] Based on these Raman signatures, we can distinguish the two different phases of bulk MoTe$_2$.

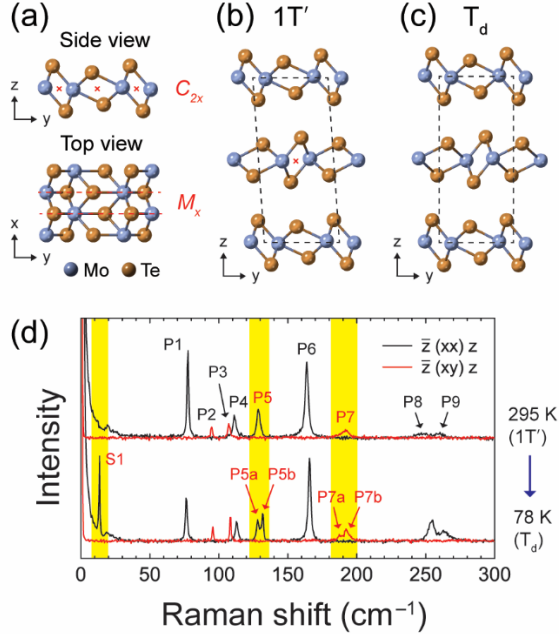

**Figure 1.** (a) Side and top view of the crystal structure of monolayer 1T′ or $T_d$ MoTe$_2$. Symmetry elements of two-fold screw axis along the $x$ direction ($C_{2x}$) and horizontal mirror plane ($M_x$) are indicated by red cross marks and dashed lines, respectively. An inversion center lies on the two-fold screw axis. (b) Side view of the crystal structure of monoclinic 1T′ MoTe$_2$. A two-fold screw axis (and thereby an inversion center) is indicated by a red cross mark. (c) Side view of the crystal structure of orthorhombic $T_d$ MoTe$_2$. (d) Polarized Raman spectra of bulk 1T′ MoTe$_2$ at 295 K and at 78 K after phase transition to the $T_d$ phase. Black (red) spectra correspond to the parallel (cross) polarization configuration with the excitation polarization along the $x$ direction. The Raman signatures of structural phase transition are highlighted in yellow.

Figures 2a and b show the two different sets of polarized Raman spectra of few-layer MoTe$_2$ with different thicknesses, measured at room temperature. Although all the samples were prepared in the same way from the same bulk 1T′ MoTe$_2$ crystal, the exfoliated few-layer samples of the same thickness show different Raman spectra. The Raman spectra of the few-layer samples exhibit the similar signatures of the 1T′ or $T_d$ phase of bulk crystals, and the samples were classified according to the line shape of P5: those with a split peak of P5a and P5b were identified as the $T_d$ phase, and those with a single peak as the 1T′ phase. For the monolayer (1L) case, the 1T′ and $T_d$ phases are not distinct. For bilayer (2L), the difference in the crystal structure is subtle, and the Raman spectra are almost the same. We found that some 2L samples have a slightly asymmetric line shape for P5 and identified those as being the $T_d$ phase. The two different phases of the same thickness show negligible differences in the high-frequency Raman spectra aside from P5. P1 and P2 in both the phases show pronounced thickness dependence as detailed in Figure 2c: as the thickness increases, the separation between P1 and P2 increases due to the redshift of P1 and the blueshift of P2. For few-layer $MoTe_2$, the group theoretical notations for the vibration modes are as follows: the vibration modes reduce into Raman-active A′ and A″ modes for both even-layer 1T′ and few-layer $T_d$ phases, and they are observed in the parallel [(*xx*) or (*yy*)] and cross polarization configurations [(*xy*) or (*yx*)] along the crystalline axes, respectively (for more details, see Supporting Information Note S1). For odd-layer 1T′ phase, the notations are identical to those for the bulk 1T′ phase.

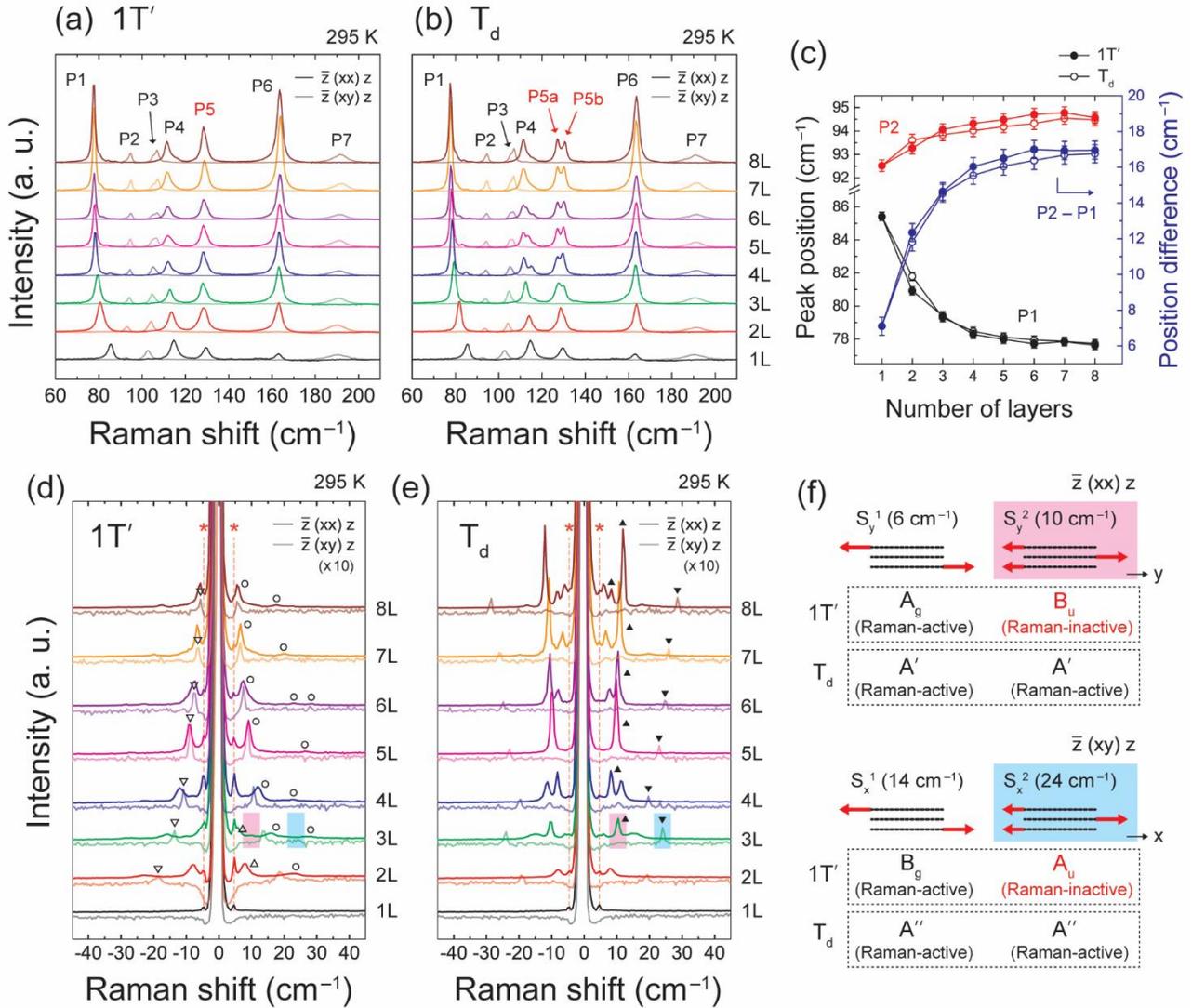

**Figure 2.** Polarized Raman spectra in the high-frequency region (60 cm$^{-1}$ to 210 cm$^{-1}$) of MoTe$_2$ in (a) 1T′ phase and (b) T$_d$ phase. Dark (light) spectra correspond to the parallel (cross) polarization configuration with the excitation polarization along the *x* direction. (c) Peak position of P1 and P2, and the difference between them as a function of thickness. Polarized Raman spectra in the low-frequency region of MoTe$_2$ in (d) 1T′ phase and (e) T$_d$ phase. Raman spectra in the cross polarization are magnified by a factor of 10. The Brillouin-scattering signal from the silicon substrate (~5 cm$^{-1}$) is marked by asterisks (*). Interlayer breathing modes (O) and shear modes in the $y$ ($\triangle$) and $x$ ($\triangledown$) directions in (d) are indicated by symbols. Extra Raman modes in the $T_d$ phase are indicated by '▲' in the parallel polarization and '▼' in the cross polarization in (e). (f) Schematic diagram of interlayer shear modes in 3L MoTe$_2$. The irreducible representation and the Raman activity of each vibration are shown for the 1T′ and $T_d$ phases. The higher-energy interlayer shear modes of 3L MoTe$_2$ allowed only in the $T_d$ phase are highlighted in the Raman spectra for comparison.

Figures 2d and e show low-frequency Raman spectra of few-layer 1T′ and $T_d$ MoTe$_2$ (see Supporting Information Figure S2 for magnified Raman spectra). Unlike high-frequency Raman modes, interlayer vibration modes are highly sensitive to the interlayer interaction and layer stacking, leading to significant differences in low-frequency Raman spectra between the two phases. For 1T′ MoTe$_2$, the breathing modes (B) and the shear modes in the $y$ direction (S$_y$) are observed in the parallel polarization configuration, and the shear modes in the $x$ direction (S$_x$) in the cross polarization configuration. B and S$_y$ modes cannot be distinguished by the polarization selection rule alone but are classified by comparing with the linear chain model calculations as will be explained. For the $T_d$ phase, several extra peaks are observed: the peaks marked by '▲' are observed in the parallel polarization configuration and assigned to the S$_y$ modes. In the cross polarization configuration, the extra peaks are marked by '▼' and assigned to the S$_x$ modes.

Since all the samples were exfoliated from a 1T′ MoTe$_2$ crystal, observation of both the 1T′ and $T_d$ phases at room temperature is intriguing. He *et al.* measured Raman spectra of mechanically exfoliated MoTe$_2$ samples capped by an hBN layer and claimed that thin samples of MoTe$_2$ below ~12 nm exist only in the $T_d$ phase.[33] They attributed the stabilization of the $T_d$ phase to charge doping in semimetallic MoTe$_2$ induced by the quantum confinement effect. On the other hand, both the 1T′ and T$_d$ phases are observed in few-layer MoTe$_2$ samples grown by chemical vapor deposition (CVD).[4,39,40] In some previous studies, the appearance of the 1T′ phase in few-layer MoTe$_2$ was ascribed to the hole doping arising from the air exposure.[32,35] In our case, however, we placed the samples into a vacuum chamber immediately after exfoliation and kept them in vacuum for all measurements, minimizing air exposure. In order to see if air exposure turns our few-layer T$_d$ samples into the 1T′ phase, we intentionally exposed a 4L MoTe$_2$ sample to air for up to 1 week but found that the sample did not change to the 1T′ phase (see Supporting Information Figure S3). Furthermore, we found some MoTe$_2$ flakes that show both the phases in the same piece (see Supporting Information Figures S4c and f). It is highly unlikely that the doping density could be significantly different within a sample. Supporting Information Figure S4c also shows that the 3L area is the T$_d$ phase, whereas thicker areas are the 1T′ phase. If doping due to air exposure is causing some regions to become the 1T′ phase, the thinner area (3L) should be affected more because the doping density is likely to be higher in thinner areas. Therefore, we believe that the appearance of the 1T′ phase in our few-layer samples is unlikely to be caused by strong hole doping. The coexistence of the two phases in exfoliated samples can be understood in terms of the small total energy difference between the two phases. The calculated energy difference between the 1T′ and T$_d$ phases is only a few meV per unit cell.[41,42] The phase transition between the two phases involves a slide along the *y* direction. As we shall see below, the interlayer shear coupling constant along the *y* direction is very small, and only a small perturbation can cause a phase transition. Since the energy difference is very small even in the bulk, it is uncertain which of the two phases is more stable at room temperature for thinner samples. As the sample becomes thinner, the surface layers (top and bottom layers) that are not bound on one side become increasingly significant, and the difference in the total energy would become even smaller for the thinner samples, leading to the observation of both phases. In our experience, for 3L samples, the two phases were found with almost equal frequency. For thicker samples, we found more 1T′ samples than the $T_d$ ones. We presume that even a small amount of strain that might be exerted during the exfoliation can cause the transition from one to the other phase.

As opposed to the isotropic 2H phase, in-plane anisotropy of the 1T′ and the $T_d$ phase gives rise to the nondegenerate interlayer shear modes and different interlayer shear coupling strengths in the *x* and *y* directions. The group theory analysis further reveals the interlayer shear modes in the *x* and *y* directions have different symmetry with each other, and they are observed in different polarization configurations. Interlayer breathing modes and shear modes in the *y* direction are mirror symmetric, while interlayer shear modes in the *x* direction (Mo-chain direction) are mirror antisymmetric. As such, interlayer breathing modes and shear modes in the *y* direction are observed in the parallel polarization, and interlayer shear modes in the *x* direction are observed in the cross polarization. In Figure 2f, the interlayer shear modes of 3L MoTe$_2$ with the corresponding irreducible representation and the Raman activity are summarized for the 1T′ and $T_d$ phases. In the 1T′ phase, only the lower-energy interlayer shear mode is Raman-active, and both the interlayer shear modes are Raman-active in the $T_d$ phase. The higher-energy interlayer shear modes at 10 cm$^{-1}$ and 24 cm$^{-1}$ of the $T_d$ phase are Raman-inactive in the 1T′ phase and thereby suppressed in Raman spectrum. Therefore, the higher-frequency shear modes are the most obvious indicators of the phase transition from 1T′ to $T_d$ phase.

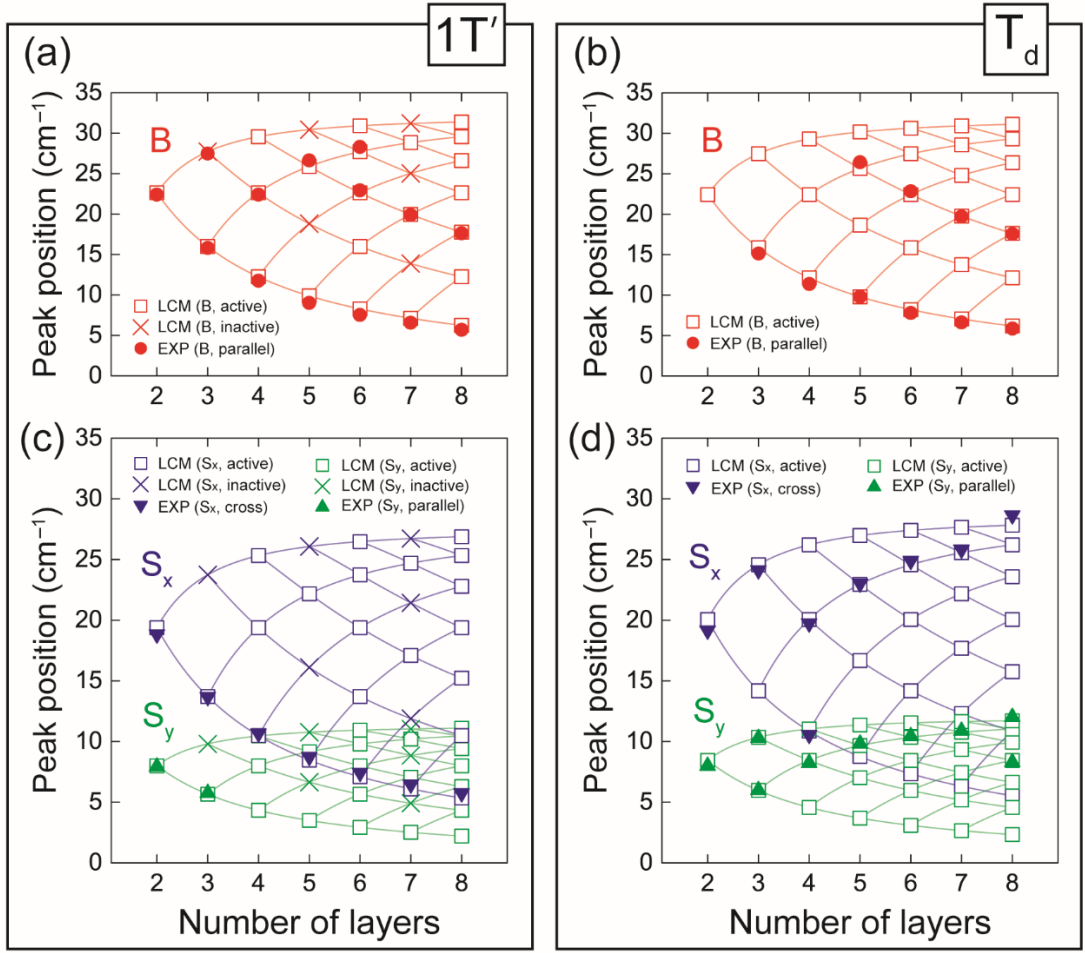

**Figure 3.** Frequency of interlayer vibration modes as a function of number of layers. Interlayer breathing (B) modes of (a) 1T′ phase and (b) $T_d$ phase, and interlayer shear modes in the $x$ ($S_x$) and $y$ ($S_y$) directions of (c) 1T′ phase and (d) $T_d$ phase. The observed Raman modes are plotted by solid symbols, and the calculated peak positions of Raman-active (Raman-inactive) modes are plotted by hollow boxes (cross marks). All the interlayer vibration modes are Raman-active in the $T_d$ phase.

Figure 3 shows the peak positions of the observed interlayer vibration modes and the calculated peak positions using the linear chain model.[43,44] The observed interlayer vibration modes are classified into interlayer breathing (B) modes and shear modes in the $x$ ($S_x$) and $y$ ($S_y$) directions from the polarization dependences and the fitting results. The extra shear modes of the $T_d$ phase (marked by '▲' and '▼' in Figure 2e) are well fitted. The interlayer coupling strengths for the two phases can be extracted from the fitting parameters and are summarized in Table 1 (for more details, see Supporting Information Note S2). Whereas the out-of-plane interlayer coupling strength is almost the same for the two phases, the in-plane interlayer coupling strength of the $T_d$ phase is slightly larger than that of the 1T′ phase (7% for the *x* direction, 11% for the *y* direction). Compared to the 2H phase,[45] the 1T′ and $T_d$ phase have much weaker interlayer coupling both in the out-of-plane and the in-plane directions. This can be explained by the considerable difference in interlayer atomic distance between tellurium atoms (Te-Te distance). The 1T′ and $T_d$ phases have some interlayer Te-Te distances that are relatively large, which possibly leads to weaker interlayer coupling compared to the 2H phase (for more details, see Supporting Information Note S3). On the other hand, the 1T′ and $T_d$ phases have similar interlayer atomic distances, and so are their interlayer coupling strengths. Furthermore, the markedly different interlayer coupling between the *x* and the *y* directions in the 1T′ and $T_d$ phases reflects the strong in-plane anisotropy, which is comparable to that of black phosphorus.[46] The phase transition between the phases is achieved by lateral sliding of the constituent layers along the *y* direction. The small interlayer coupling along this direction indicates that the required energy to slip the layers would be much smaller in the *y* direction. This is supported by the calculated energy profile, which shows the small energy difference along the interlayer shear displacements in the *y* direction.[42]

**Table 1.** Interlayer coupling strength of MoTe$_2$ in different phases

| Phase | Out-of-plane ($K_z$) (10$^{19}$ N/m$^3$) | In-plane ($K_x$) (10$^{19}$ N/m$^3$) | In-plane ($K_y$) (10$^{19}$ N/m$^3$) |
|---|---|---|---|
| 1T′ | 4.83 ± 0.08 | 3.54 ± 0.09 | 0.604 ± 0.015 |
| T$_d$ | 4.75 ± 0.10 | 3.80 ± 0.07 | 0.673 ± 0.011 |
| 2H[45] | 9.12 | 4.25 | 4.25 |

Although all the interlayer vibration modes are Raman-active in the T$_d$ phase, not all the modes are observed possibly due to small scattering cross sections. Also, the scattering cross section depends on the exact atomic positions,[47] and so the peak intensities are different in the two phases even when the modes are allowed by the Raman selection rules in both the cases. Especially, the peak intensities of the interlayer shear modes exhibit a clear difference in the two phases; the 1T′ phase has the strongest intensities for the lowest-frequency shear modes, whereas the T$_d$ phase has the strongest intensities for the higher-frequency ones. On the other hand, the peak intensities of the interlayer breathing modes are almost the same with each other. Similar trends observed in other two-dimensional materials with two different layer stacking types such as graphene (ABA and ABC stacking) or MoS$_2$ (2H and 3R stacking) were explained in terms of the stacking geometry.[47] In terms of layer stacking, the T$_d$ phase resembles the ABA-stacked graphene or 2H-stacked MoS$_2$, whereas the 1T′ phase resembles the ABC-stacked graphene or 3R-stacked MoS$_2$. The overall trends of peak intensities of interlayer vibration modes in 1T′ and T$_d$ MoTe$_2$ are similar to those of the corresponding graphene or MoS$_2$ cases, which suggests that the geometry of the layer stacking may explain the different intensities of these modes.

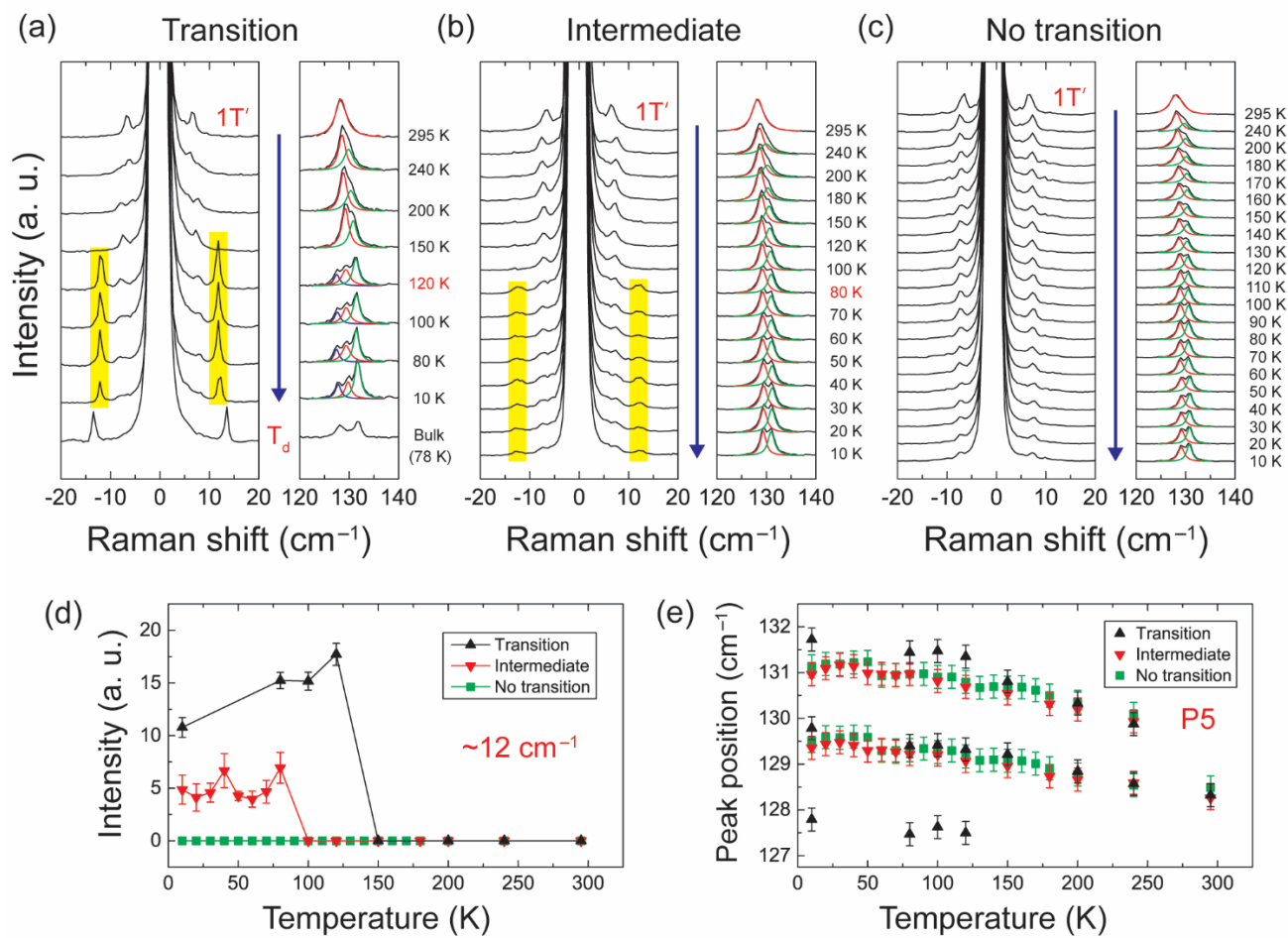

**Figure 4.** Representative Raman spectra of 7L 1T′ MoTe$_2$ measured as the temperature is lowered down to 10 K. Three different cases are shown: (a) when a phase transition to the T$_d$ phase occurs, (b) when only a slight change in interlayer vibration modes occurs (intermediate case), and (c) when no transition occurs down to 10 K. (d) Integrated intensity of an extra interlayer vibration mode at ~12 cm$^{-1}$ and (e) peak positions at ~130 cm$^{-1}$ (P5) as a function of temperature.

Based on our complete set of Raman spectra of few-layer 1T′ and T$_d$ MoTe$_2$ at room temperature, temperature-driven phase transition from the 1T′ to the T$_d$ phase was investigated by measuring the Raman spectra as the temperature was lowered to 10 K. Figure 4 shows the temperature dependence of the Raman spectra of three representative 7L 1T′ MoTe$_2$ samples (see Supporting Information Figures S7–S11 for similar data for other samples). The samples, although all were exfoliated from the same bulk MoTe$_2$ crystal, show dramatically different temperature dependences. Figure 4a is the result from a sample that shows a clear phase transition as in the case of bulk 1T′ MoTe$_2$. The phase transition occurred between 150 and 120 K accompanied by the emergence of the shear mode S$_y$ at 12 cm$^{-1}$ and the splitting of P5 into multiple peaks. On the other hand, the sample for Figure 4c shows almost no change to the low-frequency Raman spectrum although a small splitting of P5 is observed. Figure 4b is an intermediate case, in which a small signal from the extra shear mode of the T$_d$ phase is observed (shaded in yellow), but the splitting of P5 does not seem to correlate with the emergence of the extra shear mode. A structural change seems to have occurred between 100 and 80 K, although it is not obvious whether a complete phase transition has occurred. Low-frequency Raman spectra in the cross polarization configuration also show the clear features of the different transition behavior (see Supporting Information Figure S12). Figures 4d and e are the integrated intensity of the extra shear mode at ~12 cm$^{-1}$ and the positions of the different components of P5 extracted from Lorentzian fitting. From this comparison, it is evident that the most robust indicator of the phase transition is the emergence of the extra shear mode, not the splitting of P5. This interpretation is further supported by the observation that the splitting begins to occur well before the phase transition. The splitting of P5 may be attributed to Davydov splitting,[45,48,49] with the relative intensities of different Davydov-split components varying between the 1T′ and T$_d$ phases. For thinner samples, 'no transition' or 'intermediate' cases occur more commonly than the 'transition' case (see Supporting Information Figures S7–S11). Since the surface layers (top and bottom layers) constitute a larger portion for thinner samples, the difference in the total energy between the two phases would become even smaller for the thinner samples rendering the phase transition less likely. For comparison, a 3L $T_d$ MoTe$_2$ sample did not exhibit any indication of a phase transition when the temperature is lowered to 10 K (see Supporting Information Figure S13).

The high sensitivity of interlayer shear modes to the phase transition can be explained as follows: interlayer (shear) modes need to have a larger domain than intralayer modes to be well established, because the interlayer modes are defined as collective motions of atomic layers, whereas the intralayer modes involve atomic motions within each layer. The interlayer modes are also more sensitive to the layer stacking because their restoring forces rely on interlayer van der Waals interaction. Therefore, the interlayer modes are more sensitive to the phase changes than the intralayer modes.

When the temperature is raised back to 295 K, most of the samples that underwent the phase transition to the $T_d$ phase did not return to the initial 1T′ phase, except for the 35 nm and bulk samples. Even more peculiarly, some few-layer samples that did not show any transition during the cool-down to 10 K exhibited the phase transition to the $T_d$ phase or the intermediate phase during the warm-up to room temperature (see Supporting Information Figures S8b, S9c and S10a). This suggests that the $T_d$ phase probably is more stable than the 1T′ phase for few-layer MoTe$_2$.

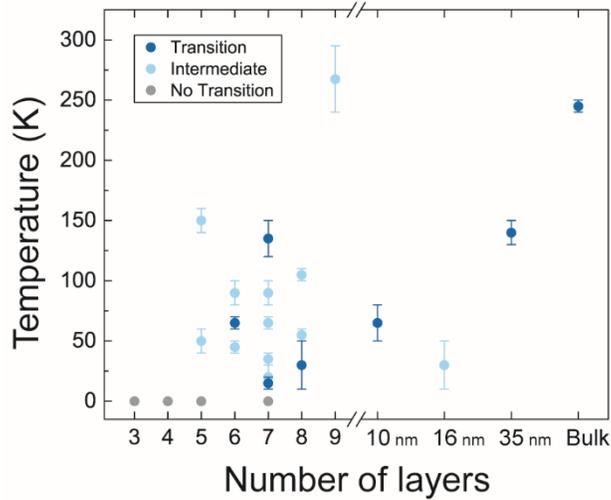

**Figure 5.** Phase transition temperature measured from several samples with different thicknesses. The critical temperatures of the transition to the $T_d$ phase and the intermediate phases are plotted by blue and cyan dots, respectively. If the transition was not observed down to 10 K, it is plotted by a gray dot at 0 K.

Figure 5 summarizes the phase transition temperatures measured during the cool-down of all the samples. For the samples that show a clear phase transition, dark symbols are used to denote the transition temperature. For the samples that show 'intermediate' transition behaviors, lighter-color symbols are used. For those samples that do not exhibit any phase transition, gray symbols are used, and the transition temperature is denoted as 0 K. We find that the phase transition temperature decreases overall, and a complete phase transition becomes rarer as the thickness decreases, and the transition is completely suppressed in 3L and 4L samples.

As we discussed earlier, the energy difference between the two phases is very small. Since the phase transition between the phases is achieved by lateral sliding of the constituent layers in the $y$ direction, an incomplete sliding of one or more layers would result in a 'mixed phase'. Indeed, such transitional phases with the mixed stacking were observed in bulk crystals.[42,50] Recently, a transitional phase ($T_d$*) with different layer stacking from 1T′ and $T_d$ phase was observed in bulk single crystals of MoTe$_2$ using elastic neutron scattering.[50] It occurred near 260 K during the warm-up from the $T_d$ phase. Although it is not obvious that this transitional phase corresponds to the 'intermediate' cases that we observed, it clearly shows the possibility of a metastable phase with different stacking structure than the 1T′ or $T_d$ phases. In fact, we found several samples that show Raman spectra that cannot be identified neither as 1T′ nor as $T_d$ phase, which might be such a metastable phase (see Supporting Information Figure S14).

One possible reason for the different phase transition behaviors exhibited by samples with the same thickness could be different charge doping levels between samples. Depending on the doping level of the starting material at room temperature, the phase transition at lower temperature can be affected. Since some phonon modes of two-dimensional materials have significant dependence on doping, the peak position or the peak width can be used to compare doping level between different samples.[51,52] For 2H MoTe$_2$,[53] for example, under electron doping of ~$10^{12}$ cm$^{-2}$, the $A_{1g}$ mode at ~170 cm$^{-1}$ shows a shift of ~1 cm$^{-1}$ and an increase in the full-width at half maximum (FWHM) of ~1 cm$^{-1}$. For 1T′ MoTe$_2$, changes in the Raman spectrum due to electron doping have been observed as well.[32] In our case, the observed Raman spectra of the 1T′ MoTe$_2$ samples that exhibit different phase transition behaviors did not show any measurable differences in the peak positions or FWHM's, which indicates that the doping level difference is not very large. Another possible origin is variations in the local strain due to corrugations on the substrate surface or unintentional strain from the sample preparation procedures. Since the 1T′ to $T_d$ phase transition is achieved by relative sliding of adjacent layers, it might be sensitive to the local strain. Some authors have ascribed the increase of the phase transition hysteresis in thinner 1T′ MoTe$_2$ to atomic defects or local strain.[34] In any case, our data show that there exist intermediate metastable phases that play increasingly more important roles in the phase transition of thinner samples. The actual phase transition path seems to depend on small sample-to-sample variations in doping or local strain. From the overall dependence of the phase transition behavior on the sample thickness, one can draw the following conclusions: the difference in the total energy between the 1T′ and the $T_d$ phases is fairly small at room temperature, resulting in coexistence of 1T′ and $T_d$ samples at room temperature. As the temperature is lowered, the total energy of the $T_d$ phase is lowered more than that of the 1T′ phase, leading to the 1T′ to $T_d$ phase transition. However, the total energy difference between the two phases at low temperatures becomes smaller as the thickness is reduced, and the phase transition often becomes incomplete due to the existence of metastable intermediate phases.

CONCLUSION

From a batch of few-layer MoTe$_2$ samples prepared from the same 1T′ bulk crystal, we identified the orthorhombic $T_d$ phase as well as the monoclinic 1T′ phase by using polarized Raman spectroscopy. The two phases exhibited a pronounced difference especially in interlayer shear modes. The observed interlayer breathing modes and shear modes were analyzed by using the linear chain model, and the extracted interlayer coupling strengths showed pronounced in-plane anisotropy, comparable to black phosphorus. Temperature-dependent Raman measurements were performed on few-layer 1T′ MoTe$_2$ samples, and the phase transition to the $T_d$ phase was observed with strong enhancement of an interlayer vibration mode and splitting of the mode at ~130 cm$^{-1}$. The phase transition temperatures were significantly lower than that for the bulk crystal, and the 3L and 4L samples did not exhibit the transition down to 10 K. For thinner MoTe$_2$, only some samples showed a well-defined phase transition, and the fraction of such samples becomes smaller as the thickness is reduced. Our data show that there are intermediate phases that are metastable between the 1T′ and $T_d$ phases, and the exact phase transition path of a given sample depends on small variations in the as-prepared sample such as doping or local strain. Since many researchers are trying to utilize the vastly different electrical and topological properties of $MoTe_2$ in the two phases, detailed information on the transition between the two phases from our work would provide a vital input in future studies of $MoTe_2$ especially in the few-layer limit.

METHODS

**Sample preparation**

The few-layer $MoTe_2$ samples were prepared on silicon substrates covered with a 280 nm silicon dioxide layer by mechanical exfoliation from a single crystal 1T′ $MoTe_2$ (HQ Graphene). The exfoliated few-layer samples are unstable in ambient conditions, and degraded samples exhibit clear features in optical image and the Raman spectrum (see Supporting Information Figure S15–S17). To minimize degradation, the samples were placed into an optical vacuum chamber just after the exfoliation and kept in vacuum ($10^{-6}$ Torr) until all the Raman measurements were completed. Only the samples without any signs of degradation in the Raman spectrum were analyzed. The thickness of the samples was first identified by the optical contrast and later compared with atomic force microscopy (AFM) and the peak positions of interlayer vibration modes in the Raman spectrum (see Supporting Information Figure S5).

**Raman measurements**

Raman measurements were carried out in vacuum ($10^{-6}$ Torr) with the 514.5 nm (2.41 eV) line of an Ar-ion laser as the excitation source. The laser beam was focused onto the sample through a 40× objective lens (0.6 N.A.) in backscattering geometry. The laser power was kept below 100 μW to avoid possible degradation and local heating of the sample. The Raman signal was obtained by using a Jobin-Yvon Horiba iHR550 spectrometer (2400 grooves/mm), combined with a liquid-nitrogen-cooled back-illuminated charge-coupled-device (CCD) detector. To achieve an ultralow-frequency cut-off (down to 5 cm$^{-1}$) in Raman measurements, volume holographic filters (OptiGrate) were used, and the signal from the Rayleigh-scattered light was effectively eliminated. The polarizations of the incident and scattered light were controlled using a half-wave plate and a polarizer. Another half-wave plate was used to keep the polarization of the signal entering the spectrometer constant with respect to the groove direction of the grating. The temperature-dependent Raman spectroscopy down to 10 K was performed by using an optical cryostat (Oxford Microstat He2).

ASSOCIATED CONTENT

**Supporting Information**.

The Supporting Information is available free of charge at

https://pubs.acs.org/doi/10.1021/acsnano.xxxxxxx.

Raman selection rules and polarization dependence of Raman modes, magnified low-frequency Raman spectra, effect of air exposure of $T_d$ MoTe$_2$, optical images and optical contrast of few-layer MoTe$_2$ samples, interlayer coupling strength calculation and interlayer atomic distances, temperature dependence of Raman spectra of all the samples, low-frequency Raman spectra in the cross polarization configuration with different transition behaviors, polarized Raman spectra of 3L T$_d$ MoTe$_2$ at 295 K and 10 K, different types of polarized Raman spectra of 5L samples, optical images and optical contrast of air-exposed sample, polarized Raman spectra of degraded sample (PDF)

AUTHOR INFORMATION

**Corresponding Author**

*E-mail: hcheong@sogang.ac.kr

**Author Contributions**

The manuscript was written through contributions of all authors. All authors have given approval to the final version of the manuscript.

**Notes**

The authors declare no competing financial interest.

ACKNOWLEDGMENT

This work was supported by the National Research Foundation (NRF) grant funded by the Korean government (MSIT) (2019R1A2C3006189 and 2017R1A5A1014862, SRC program: vdWMRC center).

For Table of Contents Only

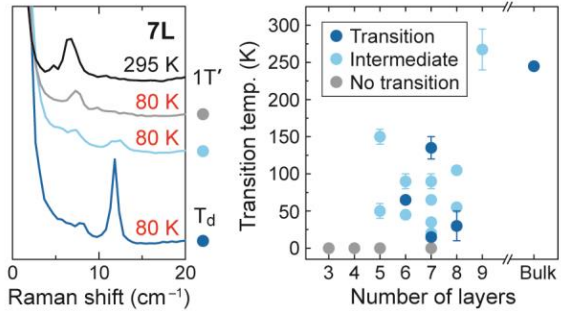

# Supporting Information

# Structural Phase Transition and Interlayer Coupling in Few-Layer 1T′ and T$_d$ MoTe$_2$

*Yeryun Cheon, Soo Yeon Lim, Kangwon Kim and Hyeonsik Cheong**

Department of Physics, Sogang University, Seoul 04107, Korea

*E-mail: hcheong@sogang.ac.kr

**Note S1. Raman selection rules**

**Figure S1. Polarization dependence of Raman modes**

**Figure S2. Magnified low-frequency Raman spectra**

**Figure S3. Effect of air exposure of T$_d$ MoTe$_2$**

**Figure S4. Optical images of few-layer MoTe$_2$ samples**

**Figure S5. Optical contrast of few-layer MoTe$_2$ samples**

**Note S2. Interlayer coupling strength calculation**

**Note S3. Interlayer atomic distances**

**Figure S6. Crystal structures of 2H and 1T′ MoTe$_2$**

**Figures S7 – S11. Temperature dependence of Raman spectra of all the samples**

**Figure S12. Low-frequency Raman spectra in the cross polarization configuration with different transition behaviors**

**Figure S13. Polarized Raman spectra of 3L T$_d$ MoTe$_2$ at 295 K and 10 K**

**Note S1. Raman selection rules**

① Bulk and odd-layer 1T′ MoTe$_2$ with the point group $C_{2h}$

The Raman tensors of A$_g$ and B$_g$ modes can be written as

$$R(A_g) = \begin{pmatrix} |a|e^{i\phi_a} & 0 & |d|e^{i\phi_d} \\ 0 & |b|e^{i\phi_b} & 0 \\ |d|e^{i\phi_d} & 0 & |c|e^{i\phi_c} \end{pmatrix} \text{ and } R(B_g) = \begin{pmatrix} 0 & |e|e^{i\phi_e} & 0 \\ |e|e^{i\phi_e} & 0 & |f|e^{i\phi_f} \\ 0 & |f|e^{i\phi_f} & 0 \end{pmatrix}. \quad (S1)$$

In the backscattering geometry with the parallel polarization configuration, the Raman intensity is given by

$$I(A_g) \propto \left| (\cos\theta \; \sin\theta \; 0) \begin{pmatrix} |a|e^{i\phi_a} & 0 & |d|e^{i\phi_d} \\ 0 & |b|e^{i\phi_b} & 0 \\ |d|e^{i\phi_d} & 0 & |c|e^{i\phi_c} \end{pmatrix} \begin{pmatrix} \cos\theta \\ \sin\theta \\ 0 \end{pmatrix} \right|^2 \quad (S2)$$

$$= |a|^2 \cos^4\theta + |b|^2 \sin^4\theta + \frac{|a||b|}{2}\cos(\phi_a - \phi_b)\sin^2(2\theta),$$

$$I(B_g) \propto \left| (\cos\theta \; \sin\theta \; 0) \begin{pmatrix} 0 & |e|e^{i\phi_e} & 0 \\ |e|e^{i\phi_e} & 0 & |f|e^{i\phi_f} \\ 0 & |f|e^{i\phi_f} & 0 \end{pmatrix} \begin{pmatrix} \cos\theta \\ \sin\theta \\ 0 \end{pmatrix} \right|^2 \quad (S3)$$

$$= |e|^2 \sin^2(2\theta).$$

Similarly, in the backscattering geometry with the cross polarization configuration, the Raman intensity is given by

$$I(A_g) \propto \left| (-\sin\theta \; \cos\theta \; 0) \begin{pmatrix} |a|e^{i\phi_a} & 0 & |d|e^{i\phi_d} \\ 0 & |b|e^{i\phi_b} & 0 \\ |d|e^{i\phi_d} & 0 & |c|e^{i\phi_c} \end{pmatrix} \begin{pmatrix} \cos\theta \\ \sin\theta \\ 0 \end{pmatrix} \right|^2 \quad (S4)$$

$$= \frac{|a|^2 + |b|^2 - 2|a||b|\cos(\phi_a - \phi_b)}{4}\sin^2(2\theta),$$

$$I(B_g) \propto \left| (-\sin\theta \quad \cos\theta \quad 0) \begin{pmatrix} 0 & |e|e^{i\phi_e} & 0 \\ |e|e^{i\phi_e} & 0 & |f|e^{i\phi_f} \\ 0 & |f|e^{i\phi_f} & 0 \end{pmatrix} \begin{pmatrix} \cos\theta \\ \sin\theta \\ 0 \end{pmatrix} \right|^2 \quad (S5)$$

$$= |e|^2 \cos^2(2\theta).$$

Therefore, only the $A_g$ ($B_g$) modes are observed in the parallel (cross) polarization configuration along the crystalline axes ($\theta = 0°$ for the $x$ direction, $\theta = 90°$ for the $y$ direction).

② Bulk $T_d$ MoTe$_2$ with the point group $C_{2v}$

The Raman tensors of $A_1$ and $A_2$ modes can be written as

$$R(A_1) = \begin{pmatrix} |a|e^{i\phi_a} & 0 & 0 \\ 0 & |b|e^{i\phi_b} & 0 \\ 0 & 0 & |c|e^{i\phi_c} \end{pmatrix} \text{ and } R(A_2) = \begin{pmatrix} 0 & |e|e^{i\phi_e} & 0 \\ |e|e^{i\phi_e} & 0 & 0 \\ 0 & 0 & 0 \end{pmatrix}. \quad (S6)$$

The Raman tensors of $B_1$ and $B_2$ modes can be written as

$$R(B_1) = \begin{pmatrix} 0 & 0 & |d|e^{i\phi_d} \\ 0 & 0 & 0 \\ |d|e^{i\phi_d} & 0 & 0 \end{pmatrix} \text{ and } R(B_2) = \begin{pmatrix} 0 & 0 & 0 \\ 0 & 0 & |f|e^{i\phi_f} \\ 0 & |f|e^{i\phi_f} & 0 \end{pmatrix}. \quad (S7)$$

Even though all the vibration modes of $A_1$, $A_2$, $B_1$ and $B_2$ symmetry are Raman active, only the $A_1$ and $A_2$ modes can be observed in the backscattering geometry.

In the backscattering geometry with the parallel polarization configuration, the Raman intensity is given by

$$I(A_1) \propto \left| (\cos\theta \quad \sin\theta \quad 0) \begin{pmatrix} |a|e^{i\phi_a} & 0 & 0 \\ 0 & |b|e^{i\phi_b} & 0 \\ 0 & 0 & |c|e^{i\phi_c} \end{pmatrix} \begin{pmatrix} \cos\theta \\ \sin\theta \\ 0 \end{pmatrix} \right|^2 \quad (S8)$$

$$= |a|^2 \cos^4\theta + |b|^2 \sin^4\theta + \frac{|a||b|}{2} \cos(\phi_a - \phi_b) \sin^2(2\theta),$$

$$I(A_2) \propto \left| (\cos\theta \quad \sin\theta \quad 0) \begin{pmatrix} 0 & |e|e^{i\phi_e} & 0 \\ |e|e^{i\phi_e} & 0 & 0 \\ 0 & 0 & 0 \end{pmatrix} \begin{pmatrix} \cos\theta \\ \sin\theta \\ 0 \end{pmatrix} \right|^2 \quad (S9)$$

$$= |e|^2 \sin^2(2\theta).$$

Similarly, in the backscattering geometry with the cross polarization configuration, the Raman intensity is given by

$$I(A_1) \propto \left| (-\sin\theta \quad \cos\theta \quad 0) \begin{pmatrix} |a|e^{i\phi_a} & 0 & 0 \\ 0 & |b|e^{i\phi_b} & 0 \\ 0 & 0 & |c|e^{i\phi_c} \end{pmatrix} \begin{pmatrix} \cos\theta \\ \sin\theta \\ 0 \end{pmatrix} \right|^2 \quad (S10)$$

$$= \frac{|a|^2 + |b|^2 - 2|a||b|\cos(\phi_a - \phi_b)}{4} \sin^2(2\theta),$$

$$I(A_2) \propto \left| (-\sin\theta \quad \cos\theta \quad 0) \begin{pmatrix} 0 & |e|e^{i\phi_e} & 0 \\ |e|e^{i\phi_e} & 0 & 0 \\ 0 & 0 & 0 \end{pmatrix} \begin{pmatrix} \cos\theta \\ \sin\theta \\ 0 \end{pmatrix} \right|^2 \quad (S11)$$

$$= |e|^2 \cos^2(2\theta).$$

Therefore, only the $A_1$ ($A_2$) modes are observed in the parallel (cross) polarization configuration along the crystalline axes ($\theta = 0°$ for the $x$ direction, $\theta = 90°$ for the $y$ direction).

③ Even-layer 1T′ MoTe$_2$ and few-layer T$_d$ MoTe$_2$ with the point group $C_s$

Raman tensors of A′ and A″ modes can be written as

$$R(A') = \begin{pmatrix} |a|e^{i\phi_a} & 0 & |d|e^{i\phi_d} \\ 0 & |b|e^{i\phi_b} & 0 \\ |d|e^{i\phi_d} & 0 & |c|e^{i\phi_c} \end{pmatrix} \text{ and } R(A'') = \begin{pmatrix} 0 & |e|e^{i\phi_e} & 0 \\ |e|e^{i\phi_e} & 0 & |f|e^{i\phi_f} \\ 0 & |f|e^{i\phi_f} & 0 \end{pmatrix}. \quad (S12)$$

In the backscattering geometry with the parallel polarization configuration, the Raman intensity is given by

$$I(A') \propto \left| \begin{pmatrix} \cos\theta & \sin\theta & 0 \end{pmatrix} \begin{pmatrix} |a|e^{i\phi_a} & 0 & |d|e^{i\phi_d} \\ 0 & |b|e^{i\phi_b} & 0 \\ |d|e^{i\phi_d} & 0 & |c|e^{i\phi_c} \end{pmatrix} \begin{pmatrix} \cos\theta \\ \sin\theta \\ 0 \end{pmatrix} \right|^2 \quad (S13)$$

$$= |a|^2 \cos^4\theta + |b|^2 \sin^4\theta + \frac{|a||b|}{2}\cos(\phi_a - \phi_b)\sin^2(2\theta),$$

$$I(A'') \propto \left| \begin{pmatrix} \cos\theta & \sin\theta & 0 \end{pmatrix} \begin{pmatrix} 0 & |e|e^{i\phi_e} & 0 \\ |e|e^{i\phi_e} & 0 & |f|e^{i\phi_f} \\ 0 & |f|e^{i\phi_f} & 0 \end{pmatrix} \begin{pmatrix} \cos\theta \\ \sin\theta \\ 0 \end{pmatrix} \right|^2 \quad (S14)$$

$$= |e|^2 \sin^2(2\theta).$$

Similarly, in the backscattering geometry with the cross polarization configuration, the Raman intensity is given by

$$I(A') \propto \left| \begin{pmatrix} -\sin\theta & \cos\theta & 0 \end{pmatrix} \begin{pmatrix} |a|e^{i\phi_a} & 0 & |d|e^{i\phi_d} \\ 0 & |b|e^{i\phi_b} & 0 \\ |d|e^{i\phi_d} & 0 & |c|e^{i\phi_c} \end{pmatrix} \begin{pmatrix} \cos\theta \\ \sin\theta \\ 0 \end{pmatrix} \right|^2 \quad (S15)$$

$$= \frac{|a|^2 + |b|^2 - 2|a||b|\cos(\phi_a - \phi_b)}{4}\sin^2(2\theta),$$

$$I(A'') \propto \left| \begin{pmatrix} -\sin\theta & \cos\theta & 0 \end{pmatrix} \begin{pmatrix} 0 & |e|e^{i\phi_e} & 0 \\ |e|e^{i\phi_e} & 0 & |f|e^{i\phi_f} \\ 0 & |f|e^{i\phi_f} & 0 \end{pmatrix} \begin{pmatrix} \cos\theta \\ \sin\theta \\ 0 \end{pmatrix} \right|^2 \quad (S16)$$

$$= |e|^2 \cos^2(2\theta).$$

Therefore, only the A' (A'') modes are observed in the parallel (cross) polarization configuration along the crystalline axes ($\theta = 0°$ for the $x$ direction, $\theta = 90°$ for the $y$ direction).

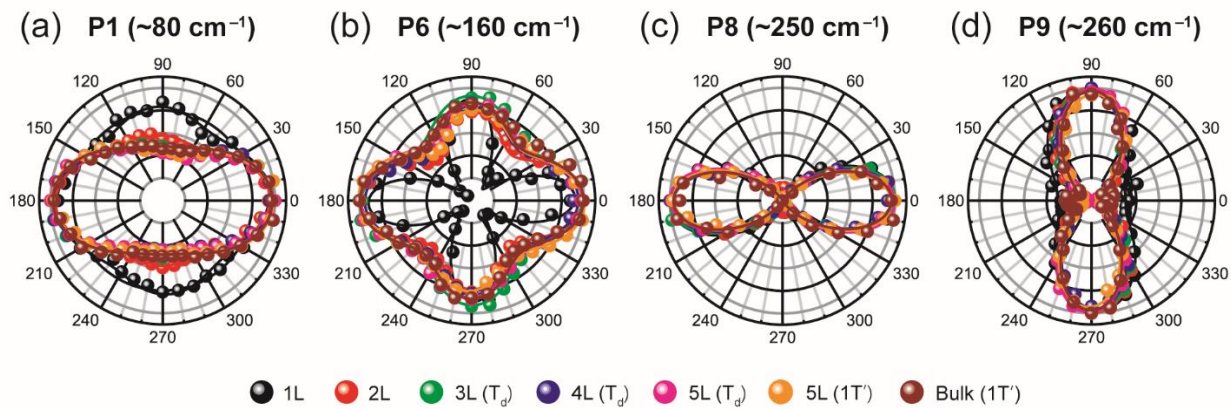

**Figure S1.** Polarization dependence of some representative Raman peaks measured in parallel polarization configuration. In the polar plot, the *x* axis corresponds to 0°. The solid curves are fits to the calculated Raman intensity. The crystalline orientation can be determined from the polarization dependence of the Raman modes.[S1–S3]

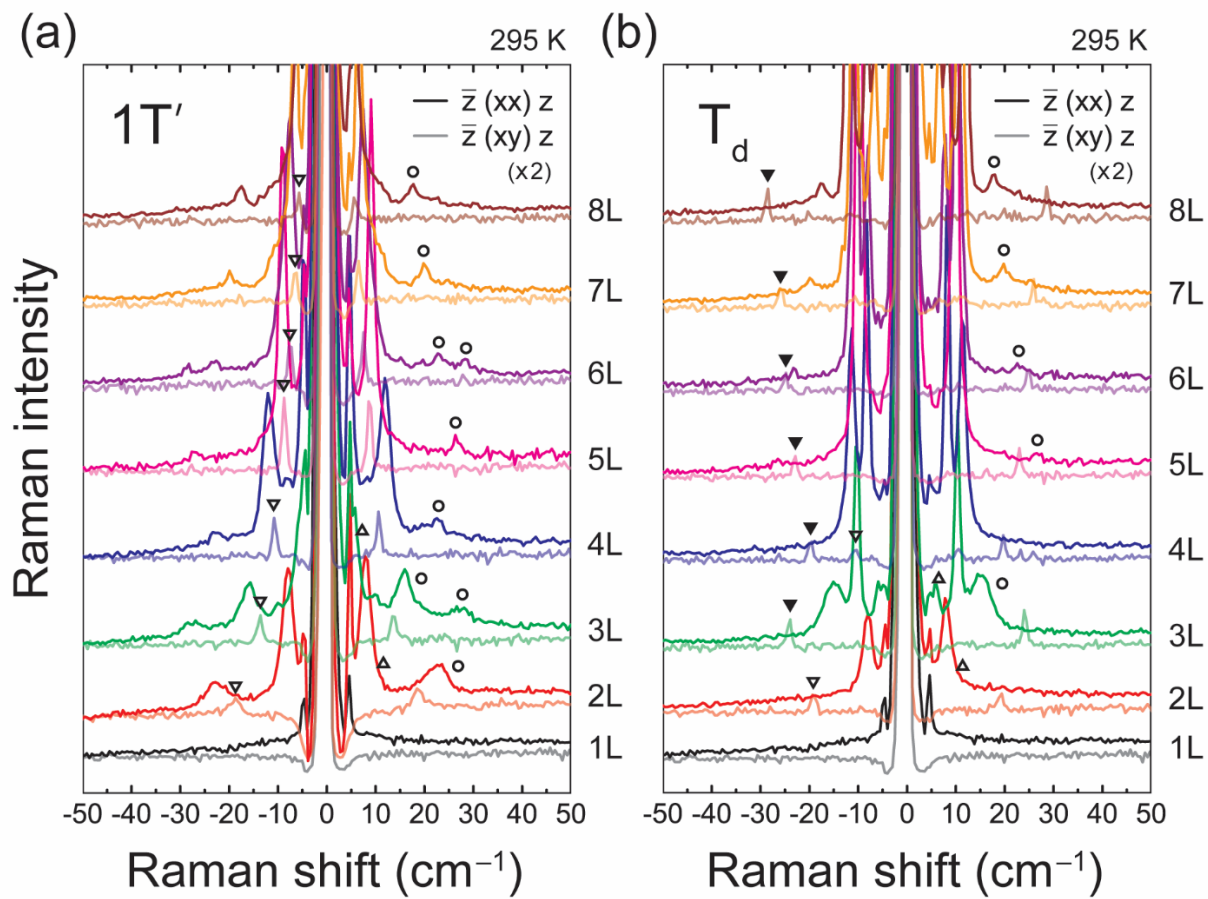

**Figure S2.** Magnified low-frequency Raman spectra of 1T′ and T$_d$ MoTe$_2$. Raman spectra in the cross polarization are magnified by a factor of 2. Interlayer breathing modes of the 1T′ and T$_d$ phases are indicated by 'O', and interlayer shear modes in the $y$ and $x$ directions of the 1T′ phase are indicated by '△' and '▽', respectively. Extra shear modes in the $x$ direction of the T$_d$ phase are indicated by '▼'.

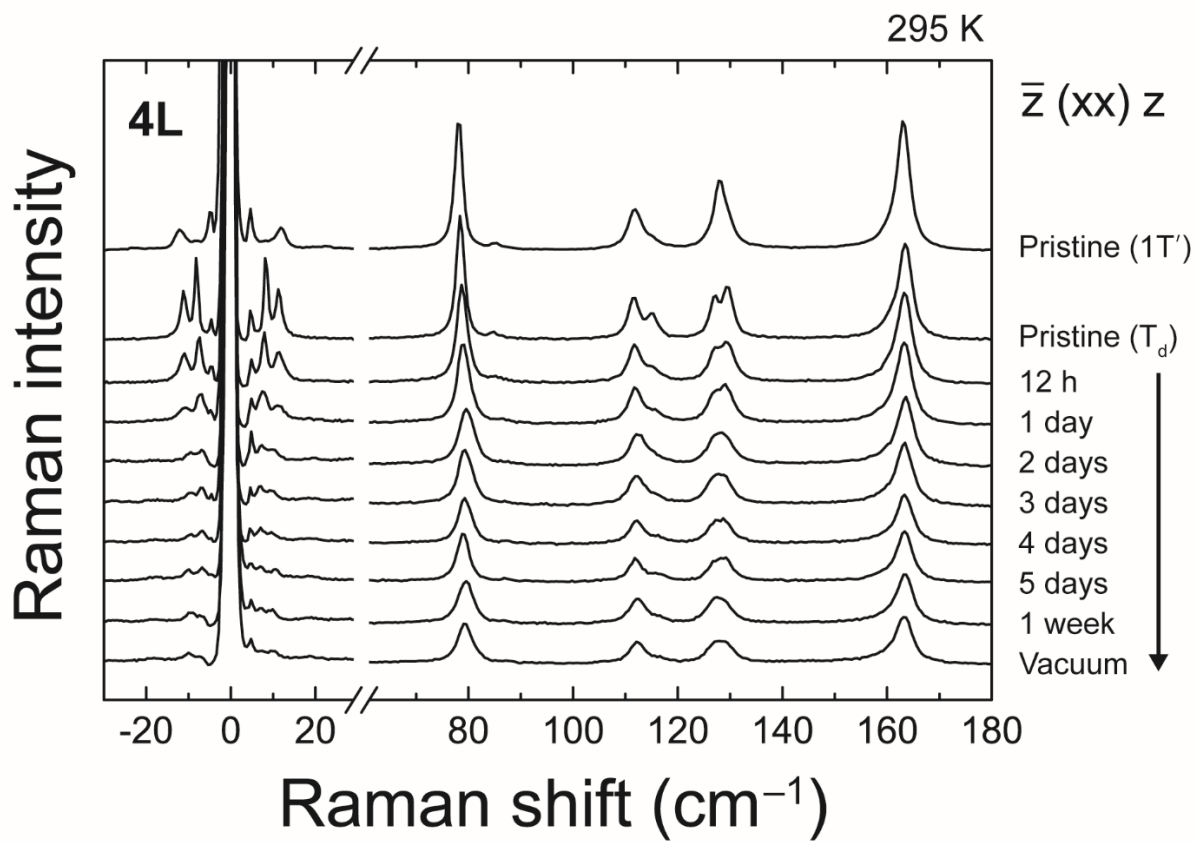

**Figure S3.** Effect of air exposure of $T_d$ MoTe$_2$. Polarized Raman spectra of 4L $T_d$ MoTe$_2$ are measured with varying air exposure. Upon air exposure, the Raman peaks at ~130 cm$^{-1}$ become broad, and the peak splitting is not resolved. Also, the intensity of the low-frequency Raman modes significantly decreases. When the sample was put into vacuum after air exposure, the Raman spectrum does not revert to that of pristine $T_d$ MoTe$_2$, indicating that permanent degradation had occurred. Also, a Raman spectrum of a pristine 4L 1T′ MoTe$_2$ is shown for comparison. It is clear that the air exposure did not change the $T_d$ MoTe$_2$ sample to 1T′ MoTe$_2$.

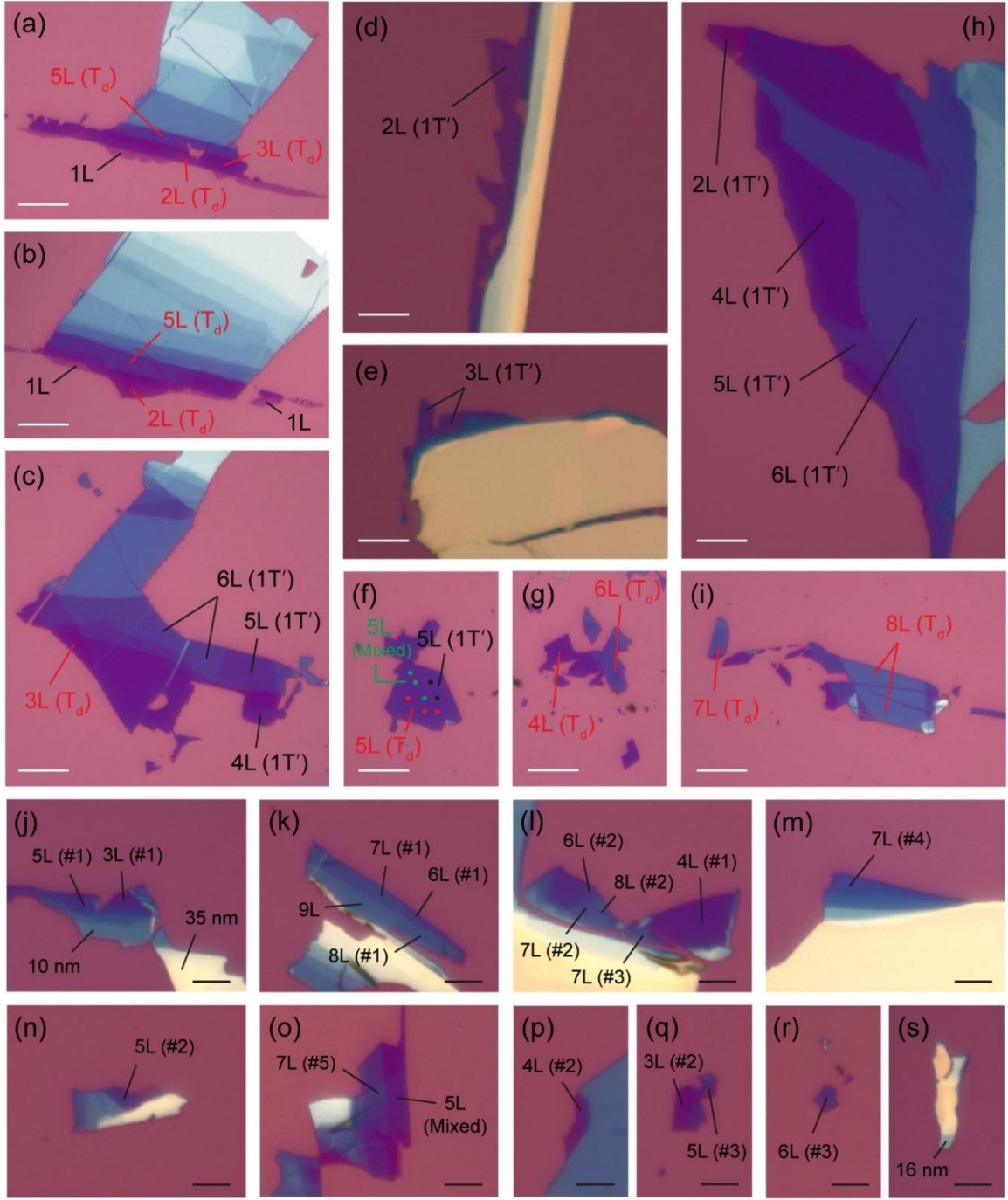

**Figure S4.** Optical images of few-layer MoTe$_2$ samples in different phases. Samples shown in (j) – (s) are 1T′ MoTe$_2$ (at room temperature) samples used for temperature-dependent Raman measurements. The sample numbers match with those in Figures S7–S12. The white and black scale bars are 20 μm and 10 μm, respectively.

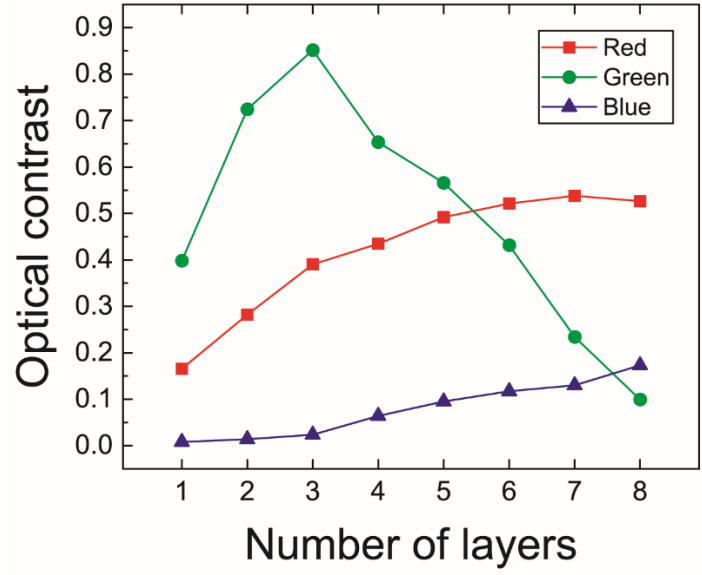

**Figure S5.** Optical contrast of red, green and blue as a function of number of layers. The optical contrast is extracted by using the following definition: $\text{O.C.} = \dfrac{|\,I(\text{sample}) - I(\text{substrate})\,|}{I(\text{substrate})}$ .

**Note S2. Interlayer coupling strength calculation**

In the linear chain model, assuming only the nearest-neighbor interactions, the Lagrangian of $N$ particles of mass $m$ and displacement $x_j$ from the equilibrium can be written as

$$L = \frac{1}{2} \sum_{j=1}^{N+1} \left[ m \dot{x}_j^2 - k(x_{j-1} - x_j)^2 \right], \tag{S17}$$

where $k$ is the force constant between the particles. By applying Lagrange's equation, one can obtain the following matrix equation:

$$m \ddot{x}_j - k(x_{j-1} - 2x_j + x_{j+1}) = 0. \tag{S18}$$

By solving the characteristic equation, one can obtain the $\alpha$-th eigenvalue $\omega_\alpha$ from suitable boundary conditions:

$$\omega_\alpha = 2 \sqrt{\frac{k}{2m} \left[ 1 - \cos\left( \frac{(\alpha-1)\pi}{N} \right) \right]}, \tag{S19}$$

where $m$ and $k$ are in SI units and the phonon frequency has the unit of [rad/s]. To express $\omega_\alpha$ in terms of the Raman shift ($1/\lambda = \omega/2\pi c$) in [cm$^{-1}$], the above expression should be divided by a factor of $2\pi c$. Then, the frequency of the $\alpha$-th interlayer vibration mode in $N$-layer MoTe$_2$ can be written as

$$\omega_\alpha = \frac{1}{\pi c} \sqrt{\frac{K}{2\mu} \left[ 1 - \cos\left( \frac{(\alpha-1)\pi}{N} \right) \right]}, \quad \text{where } \alpha = 2, 3, \ldots, N. \tag{S20}$$

Here, $c$ is the speed of light in vacuum, $\mu$ the mass per unit area, and $K$ the interlayer coupling strength (or the force constant) per unit area which is what we are looking for.[S4,S5] By using the mass per unit area of $\mu = 5.31 \times 10^{-6}$ kg/m$^2$ for 1T' and T$_d$ MoTe$_2$, we obtain the interlayer coupling strengths.

**Note S3. Interlayer atomic distances**

The 1T′ and 2H phases have the similar c-axis lattice constant, 13.86 Å for the 1T′ phase and 13.97 Å for the 2H phase at room temperature,[S6,S7] and so it seems that they have the similar interlayer distance. However, the 1T′ and $T_d$ phases have different monolayer structure from the 2H phase, and thereby the simple comparison of the overall interlayer distance hardly captures the difference in the interlayer coupling strength. Instead, we should compare the interlayer atomic distances between tellurium atoms (Te-Te distances) that would play a major role in the overall interlayer interaction. For both the 2H and 1T′ phases, each tellurium atom has three adjacent tellurium atoms in the next layer, and we have extracted the three interlayer Te-Te distances for each tellurium atom, as shown in Tables S1 and S2. For the 1T′ phase, there are 4 inequivalent Te positions as indicated in Figure S6. We have noticed that some tellurium atoms in the 1T′ phase have much larger interlayer Te-Te distances compared to the 2H phase. This is also true for the $T_d$ phase since the 1T′ and $T_d$ phases have only a slight structural difference. Therefore, the relatively large Te-Te distances in the 1T′ and $T_d$ phases would lead to weaker interlayer coupling compared to the 2H phase.

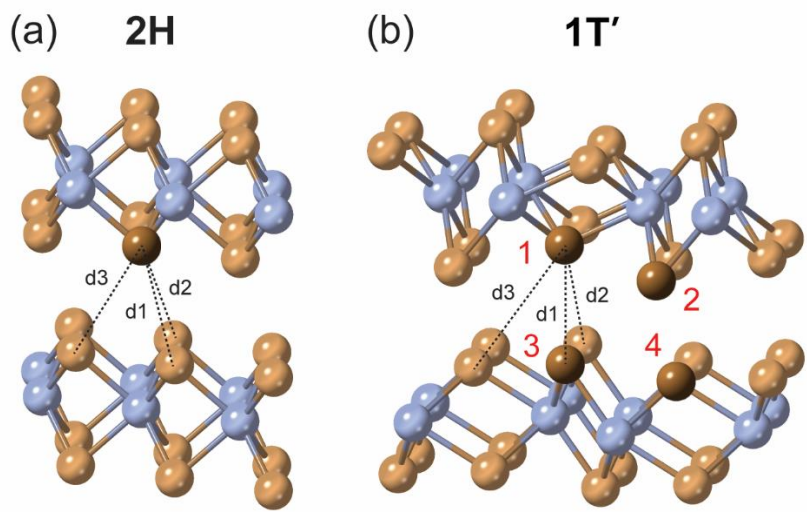

**Figure S6.** Crystal structures of 2H and 1T′ $MoTe_2$.

**Table S1.** Interlayer Te-Te distance in 2H phase (in Å)

| 2H | Te 1 |
|---|---|
| d1 | 3.945 |
| d2 | 3.945 |
| d3 | 3.945 |

**Table S2.** Interlayer Te-Te distance in 1T′ phase (in Å)

| 1T′ | Te 1 | Te 2 | Te 3 | Te 4 |
|---|---|---|---|---|
| d1 | 3.857 | 3.861 | 3.857 | 3.861 |
| d2 | 3.857 | 3.861 | 3.857 | 3.861 |
| d3 | 4.837 | 3.908 | 3.908 | 4.837 |

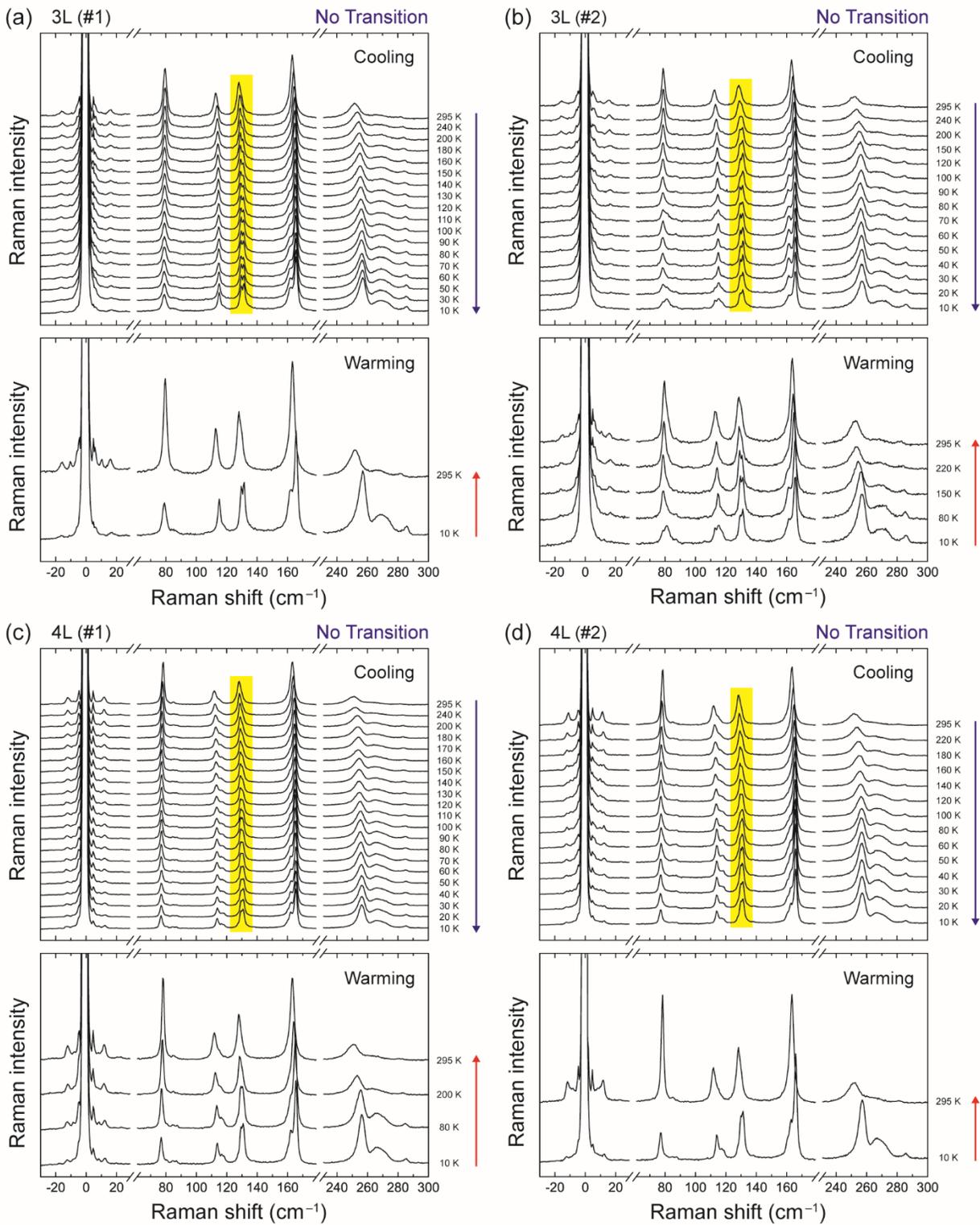

**Figure S7.** Temperature dependence of Raman spectra of 3L (#1, #2) and 4L (#1, #2) samples.

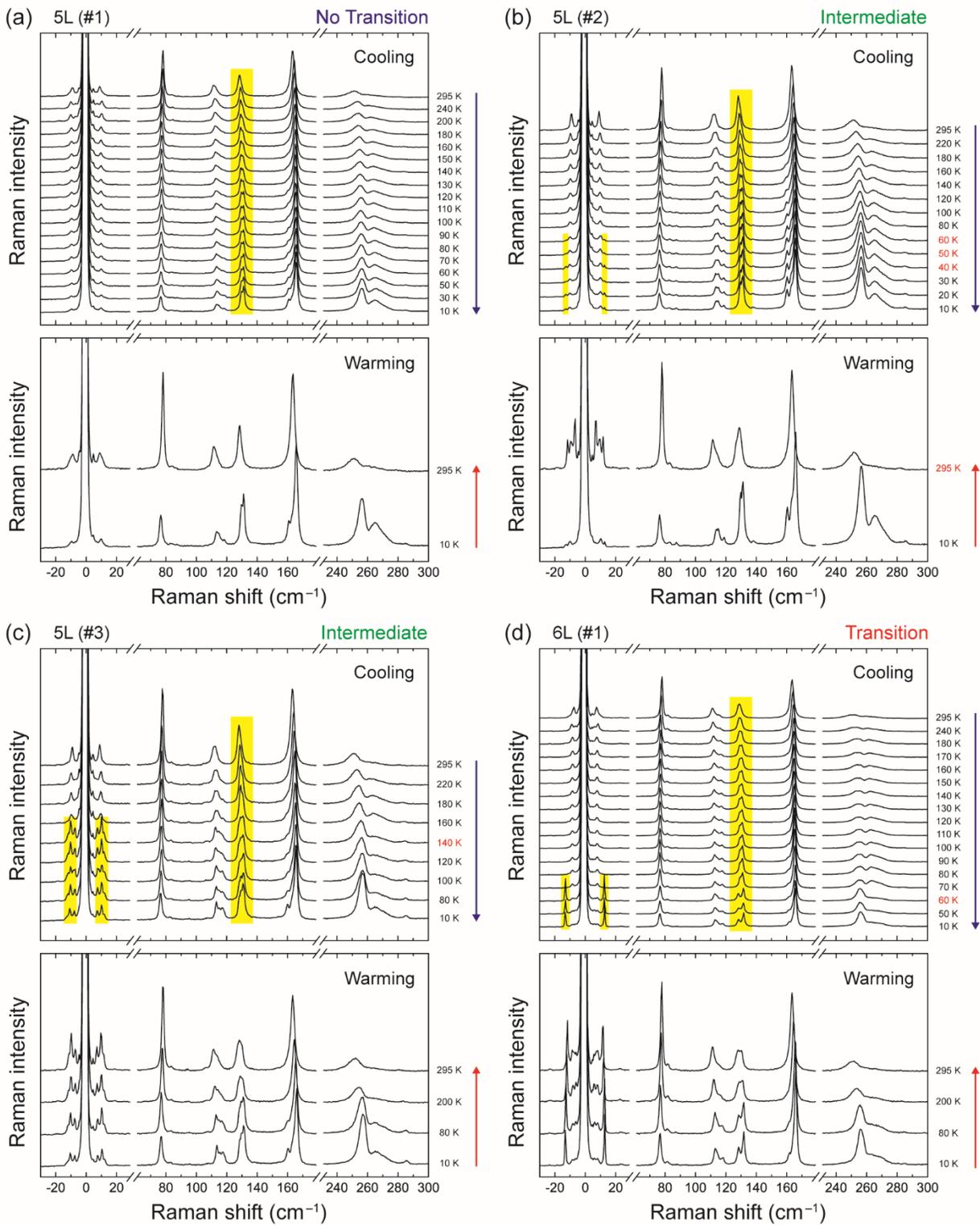

**Figure S8.** Temperature dependence of Raman spectra of 5L (#1, #2, #3) and 6L (#1) samples.

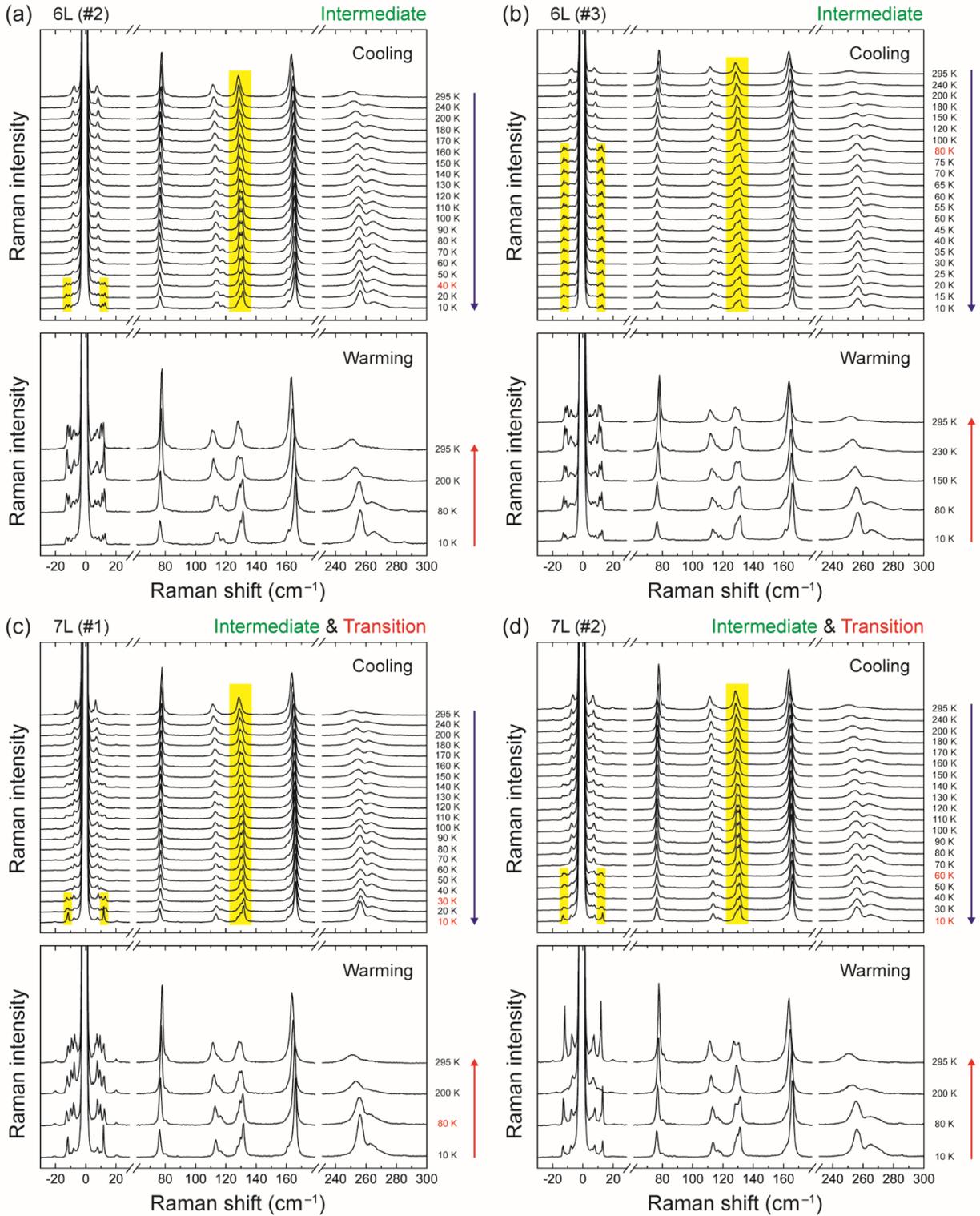

**Figure S9.** Temperature dependence of Raman spectra of 6L (#2, #3) and 7L (#1, #2) samples.

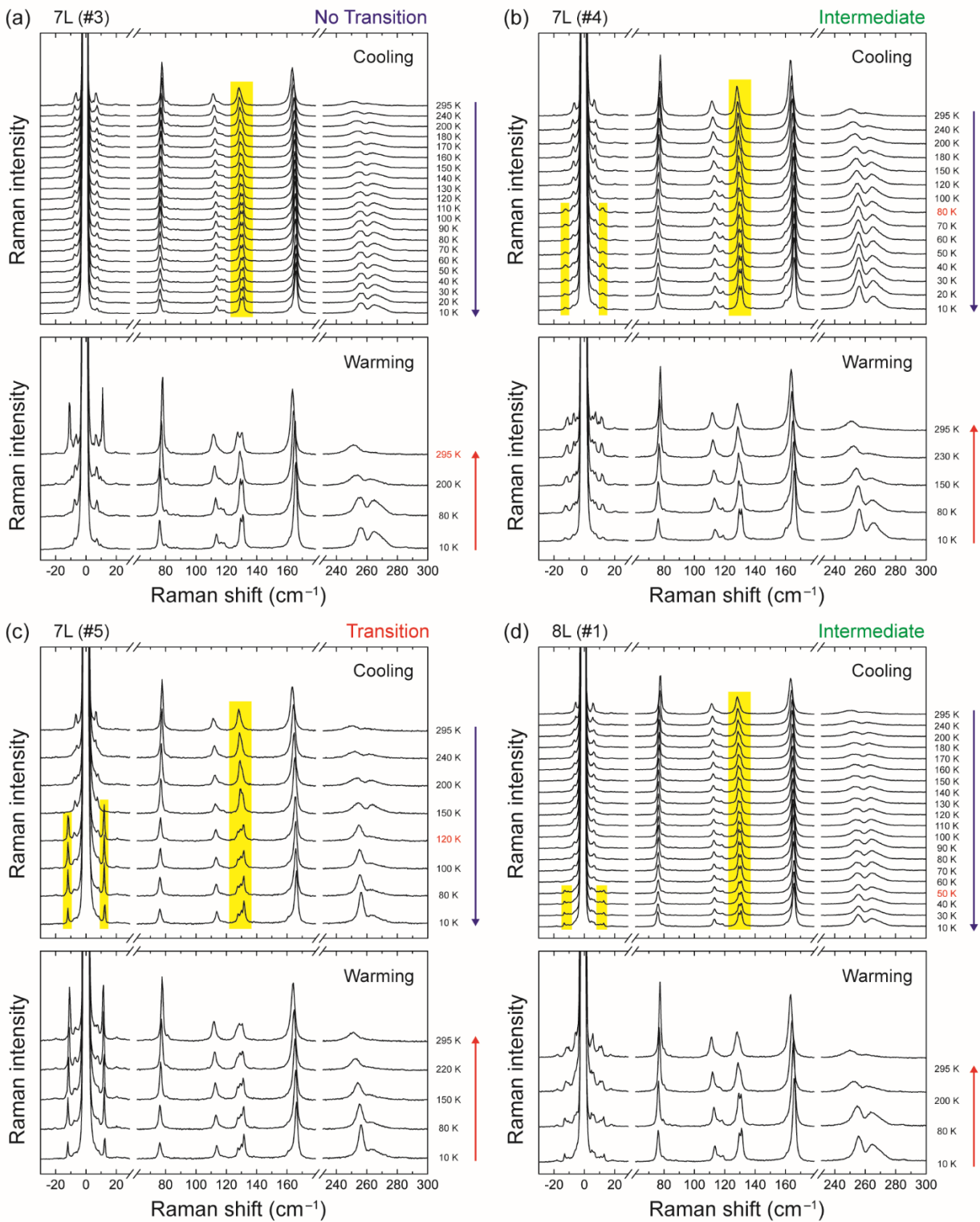

**Figure S10.** Temperature dependence of Raman spectra of 7L (#3, #4, #5) and 8L (#1) samples.

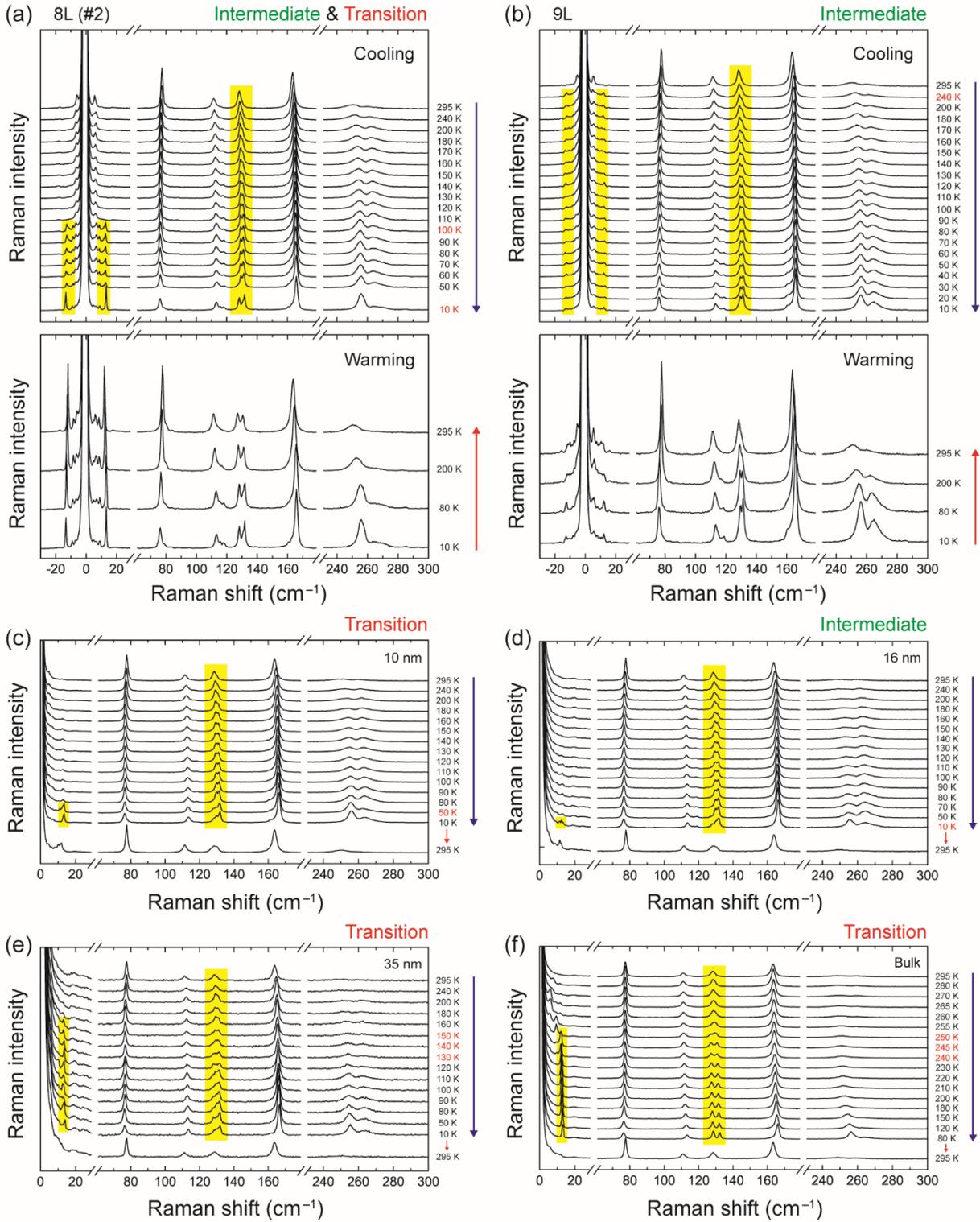

**Figure S11.** Temperature dependence of Raman spectra of 8L (#2), 9L, 10 nm, 16 nm, 35 nm and bulk samples.

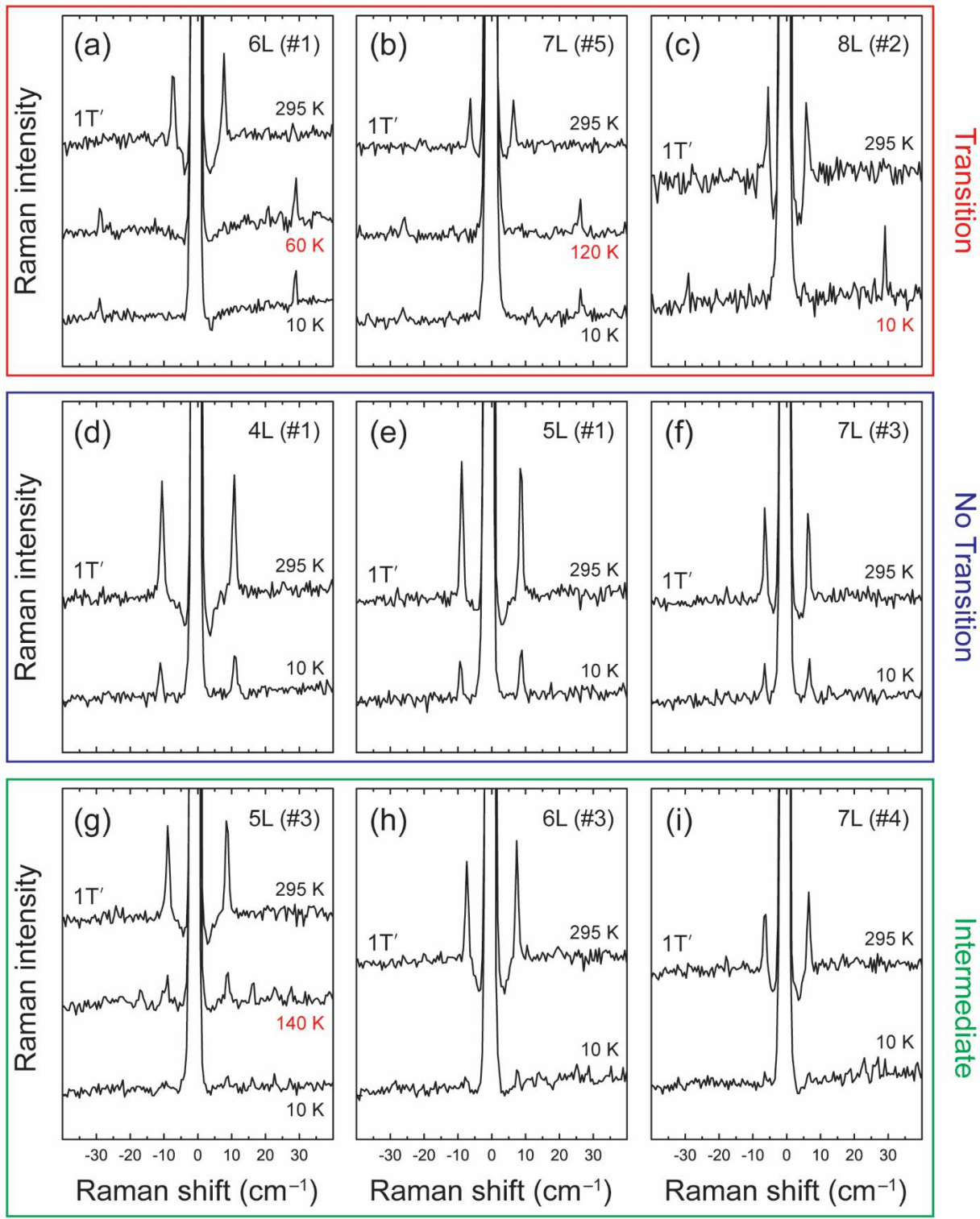

**Figure S12.** Low-frequency Raman spectra in the cross polarization configuration of a few samples that exhibit different transition behaviors.

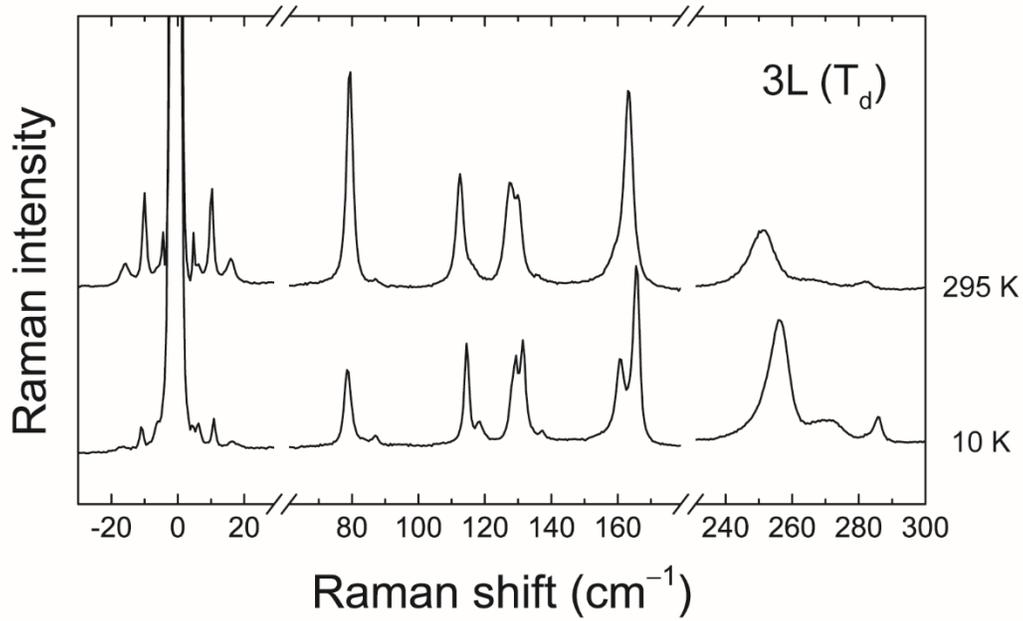

**Figure S13.** Polarized Raman spectra of 3L $T_d$ MoTe$_2$ at 295 K and 10 K. The low-frequency modes and the splitting of P5 do not show any major change (other than temperature-induced shifts) at the two temperatures, indicating the lack of a phase transition.

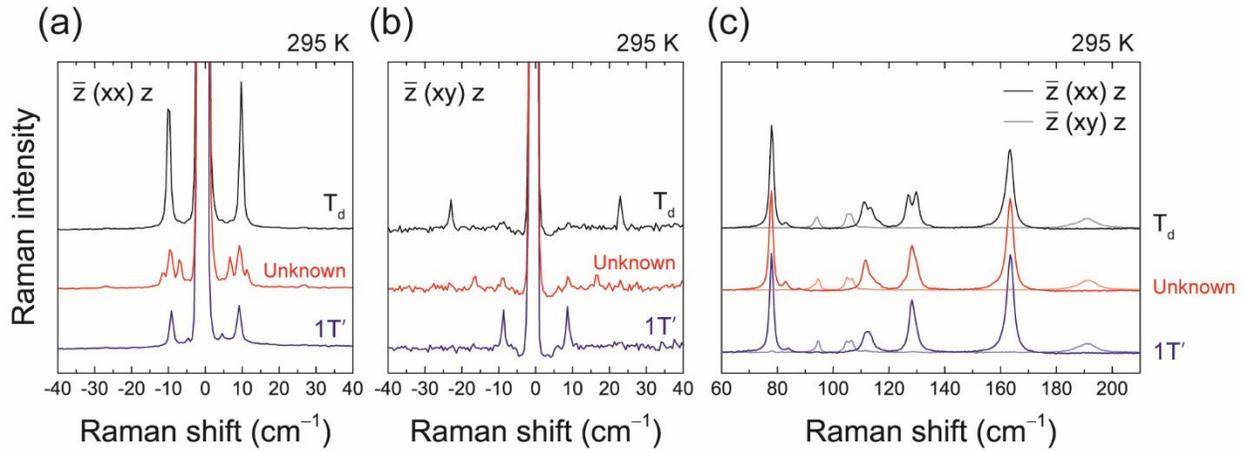

**Figure S14.** Different types of polarized Raman spectra observed in exfoliated 5L samples at room temperature. Low-frequency Raman spectra in (a) parallel and (b) cross polarization configurations. (c) High-frequency Raman spectra in parallel and cross polarization configurations. The sample that shows a Raman spectrum with an asymmetric peak shape at ~130 cm$^{-1}$ is deemed to be a mixed phase.

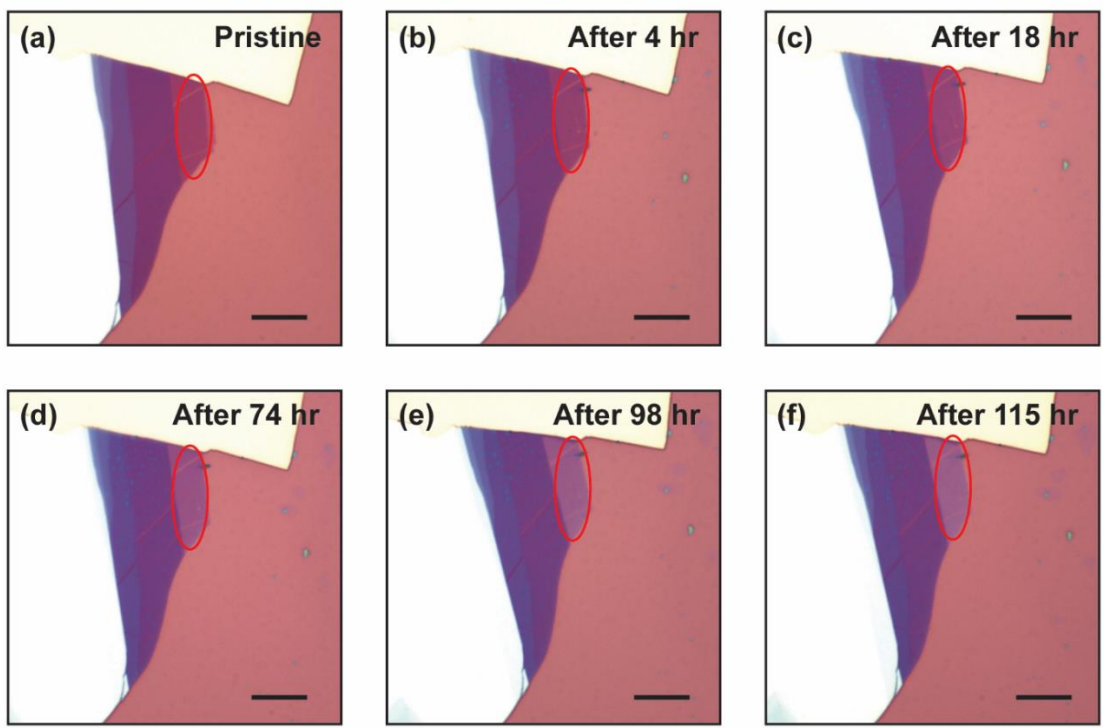

**Figure S15.** Optical images of a sample exposed to air. The 2L region indicated in red shows a significant optical-contrast change with increasing air exposure. The scale bar is 20 μm.

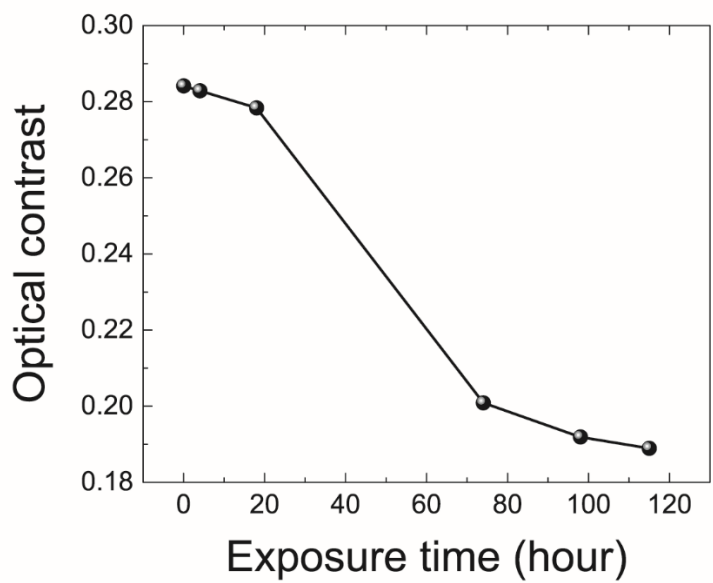

**Figure S16.** Optical contrast of the 2L region as a function of air-exposure time.

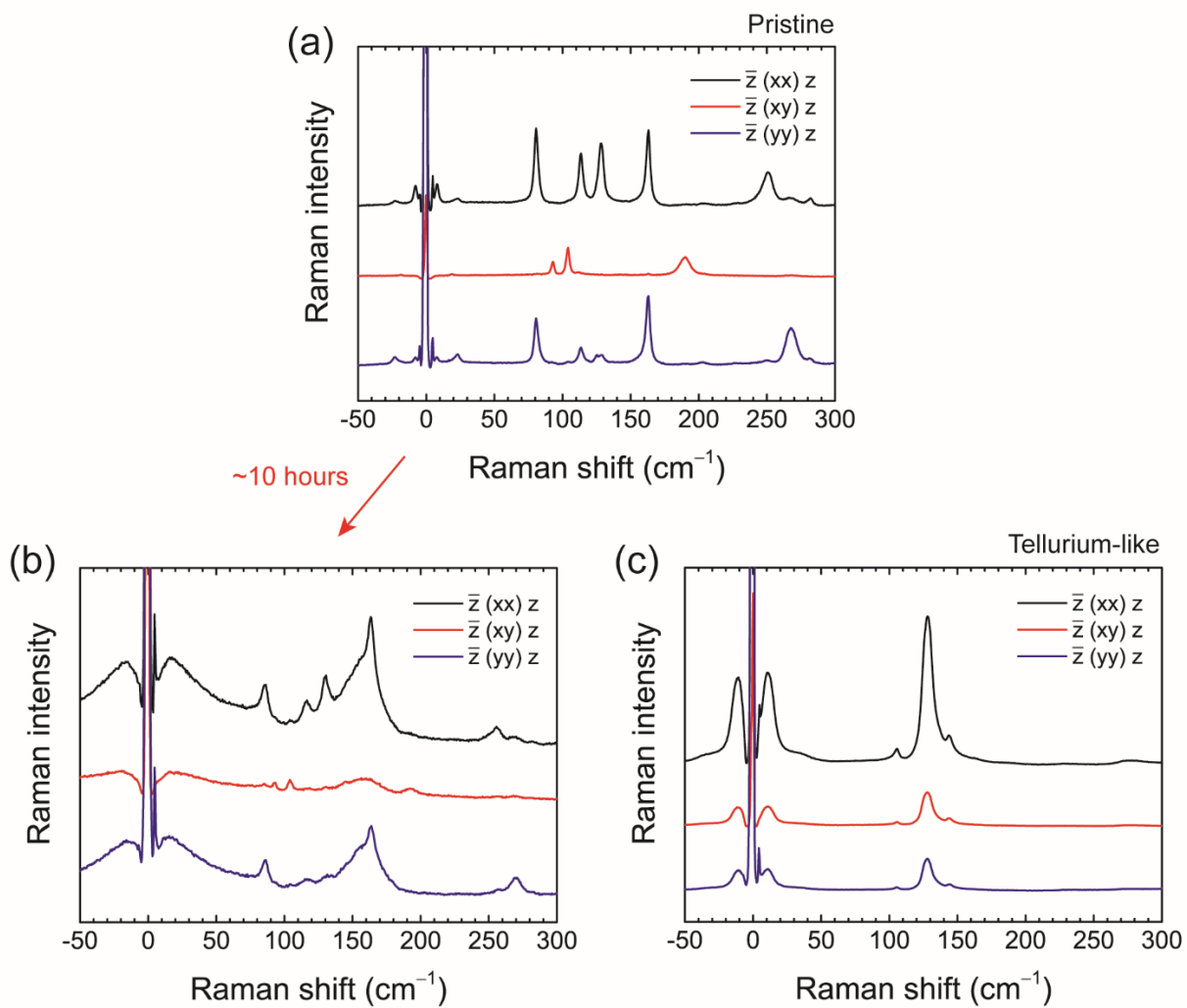

**Figure S17.** Polarized Raman spectra of (a) pristine 2L 1T′ MoTe$_2$ and (b) 2L 1T′ MoTe$_2$ exposed to ambient air for ~10 hours. After air exposure, the intensity of Raman modes decreases, and the peaks broaden significantly. (c) Tellurium-like Raman spectrum observed in some MoTe$_2$ samples after air exposure. Tellurium clusters are formed possibly due to the weak bonding between molybdenum and tellurium.[S8]